\documentclass[preprint,aps,showpacs,floatfix]{revtex4}
\usepackage{amsmath,amssymb,graphicx,epsfig}

\newcommand{\vv}{{\bf {v}}}
\newcommand{\rr}{{\bf {r}}}
\newcommand{\pp}{{\bf {p}}}
\newcommand{\qq}{{\bf {q}}}

\newcommand{\ep}{{\mbox{$\epsilon_p$}}}
\newcommand{\eF}{{\mbox{$\epsilon_F$}}}
\newcommand{\vF}{{\mbox{$v_F$}}}
\newcommand{\pF}{{\mbox{$p_F$}}}

\begin{document}
%\draft
%\def\pd#1#2{\frac{\partial #1}{\partial #2}}

\title{\bf Interaction of ballistic quasiparticles and
vortex configurations in superfluid $^3$He-B}

\author{C. F. Barenghi${}^1$, Y. A. Sergeev${}^2$, N. Suramlishvili${}^{1,3}$, and P. J. van Dijk${}^2$}

\affiliation {
${}^1$School of Mathematics and Statistics,
Newcastle University, Newcastle upon Tyne, NE1 7RU,
${}^2$School of Mechanical and Systems Engineering,
Newcastle University, Newcastle upon Tyne, NE1 7RU,
${}^3$Andronikashvili Institute of Physics, Tbilisi,
0177, Georgia}
\date {\today}

\begin {abstract}
The vortex line density of turbulent superfluid $^3$He-B at very low
temperature is deduced by detecting the shadow of ballistic
quasiparticles which are Andreev reflected by quantized vortices.
Until now the measured total shadow has been interpreted as the sum
of shadows arising from interactions of a single quasiparticle with
a single vortex. By integrating numerically the quasi-classical
Hamiltonian equations of motion of ballistic quasiparticles in the
presence of nontrivial but relatively simple vortex systems (such as
vortex-vortex and vortex-antivortex pairs and small clusters of
vortices) we show that partial screening can take place, and the
total shadow is not necessarily the sum of the shadows. We have also
found that it is possible that, upon impinging on complex vortex
configurations, quasiparticles experience multiple reflections,
which can be classical, Andreev, or both.
\end{abstract}

\maketitle

\section{Introduction} \label{Introduction}

The context of this work is quantum turbulence \cite{BDV2001} at
temperatures $T \ll T_c$, where $T_c$ is the critical temperature.
In this regime, the viscous normal fluid component and the mutual
friction can be neglected, and quantum turbulence takes its purest
form: a tangle of quantized vortex filaments which move in a fluid
without viscosity.

Experiments at these very low temperatures have
produced intriguing results in both $^3$He and $^4$He. In $^4$He,
McClintock and collaborators discovered that quantum turbulence, initially
generated by a moving grid, quickly decays, despite the absence of
viscous dissipation \cite{Davis2000}. In $^3$He-B, Fisher and
collaborators found that quantum turbulence,
initially confined in a small region, spreads in space and
decays \cite{Fisher2001,evaporation}.
These and other results raise  challenging questions
to low temperature physicists and fluid mechanicists alike.

In the case of homogeneous quantum turbulence, the turbulence's
intensity is characterized, at least in the first approximation, by
the vortex line density $L$ (vortex length per unit volume), a
quantity which can be measured using techniques such as second sound
and ion trapping. From the vortex line density, the typical distance
between vortices, $\ell \sim L^{-1/2}$, can be inferred. The current
understanding of quantum turbulence \cite{Vinen-Niemela} at very low
temperatures is that, at length scales much larger than $\ell$, the
nonlinear interaction between the vortex lines results in partial
alignment and polarization, such that, for $k \ll 1/\ell$, the
superfluid supports an energy cascade from large scales to small
scales, which manifests itself in the classical Kolmogorov energy
spectrum  $E_k \sim k^{-5/3}$ where $k$ is the wavenumber. Numerical
simulations performed using the vortex filament model
\cite{Araki2002} and the nonlinear Schroedinger equation model
\cite{Nore1997, Kobayashi2005} confirmed the existence of such
spectrum. The energy cascade implies the existence of an energy
sink, and the natural question arises as what should be this energy
sink in the absence of viscous dissipation. The likely energy sink
is acoustic: it is thought that kinetic energy decreases due to the
emission of phonons by Kelvin waves \cite{Vinen2001,Mitani}. Kelvin
waves are helical displacements of vortex filaments which rotate
with angular frequency $\omega \sim k^2$. To efficiently radiate
sound, $\omega$, hence $k$, must be very large: at the length scale
of vortex separation, $\ell$, sound radiation is negligible.  To
bridge this gap we have to appeal to the existence of a Kelvin wave
cascade process which generates smaller scales, and, in analogy to
the classical Kolmogorov cascade, shifts the energy to the required
high wavenumbers $k$. Numerical simulations revealed that vortex
reconnections decrease the kinetic energy directly
\cite{sound,pulse} and trigger the Kelvin wave cascade
\cite{Kivotides2001}

The details of this scenario still need to be properly understood.
First of all, the possibility has been raised that there is an
energy bottleneck between the Kolmogorov cascade at $k \ll 1/\ell$
and the Kelvin wave cascade at  $k \gg 1/\ell$
\cite{Lvov,Kozik2007}. Secondly, recently experiments
\cite{Walmsley,Golov2008} suggest the existence of a new form of
turbulence: a less structured, one-scale, "ultraquantum" turbulence
state (also called "Vinen" turbulence \cite{Volovik}), which decays
as $L \sim t^{-1}$, in contrast to the more structured, multi-scale
"semi-classical" quantum turbulence which decays as $L \sim
t^{-3/2}$ (consistently with the $k^{-5/3}$ energy spectrum).
Thirdly, the nature of the spectrum of $L$ is still unclear: if we
naively interpret $L$ as a measure of vorticity, the spectrum of $L$
should increase with $k$ if $E_k\sim k^{-5/3}$, but experiments show
otherwise \cite{Roche,Roche-Barenghi, Lancaster-spectrum}.

Homogeneous turbulence is clearly the most important turbulence
problem, but, as mentioned before, there are also experiments in
which turbulence is confined in a fraction of the experimental cell,
that is to say it is inhomogeneous and it can spread in space.
Examples of inhomogeneous or anisotropic turbulence are turbulence
generated by a vibrating wire \cite{Yano2007,BradleyJLTP2005}, or
grid \cite{BradleyPRL2005}, fork \cite{fork}, counterflow
\cite{Gordeev} and rotating counterflow \cite{Swanson,Tsubota} in
$^4$He, and the twisted vortex state accompanied by a moving vortex
front \cite{Eltsov2006,Eltsov2008} observed in rotating $^3$He-B.
Inhomogeneous turbulence may seem less generic than homogeneous
turbulence, but is equally worth of attention. The reason is that at
very low temperatures, in a pure superfluid, the key difference
\cite{Barenghi2008} between classical and quantum fluid behaviour
becomes more apparent: vortex reconnections are forbidden in a
classical inviscid Euler fluid, but can take place in a superfluid.

The experimental study of quantum turbulence would be greatly
facilitated if better visualization techniques were available.
Classical turbulence can be investigated using a large variety of
methods: ink, smoke, Kalliroscope flakes, hydrogen bubbles, hot wire
anemometry, laser Doppler anemometry, particle image velocimetry
(PIV), etc. On the contrary, there are few techniques available in
liquid helium; the most used are second sound and ion trapping in
$^4$He and NMR in $^3$He. A drawback of these techniques is that
they only measure quantities which are averaged over a large region,
and we know from the study of classical turbulence that it is
important to have local information about fluctuations. Fortunately
this problem has been recognized: work is in progress to build
smaller sensors, and new measurement techniques have been developed.
In $^4$He, at temperatures above $1~K$, a major breakthrough has
been the implementation of the PIV technique using micron-size
spheres made of glass and polymers \cite{VanSciver} and solid
hydrogen\cite{Lathrop}.

In the more difficult regime of very low temperatures $^3$He-B, the
Andreev reflection technique pioneered at Lancaster has been a major
advance in providing experimentalists with a tool for studying
turbulence. The technique is based on the fact that the dispersion
curve $E=E({\bf p})$ of quasiparticles is tied to the reference
frame of the superfluid, so, in a superfluid moving with velocity
${\bf v}_s$, the dispersion curve becomes $E({\bf p})+ {\bf p} \cdot
{\bf v}_s$, where ${\bf p}$ is the momentum
\cite{Bradley-PRL2004,Fisher2008}. Thus a side of a vortex line
presents a potential barrier to oncoming quasiparticles, which can
be reflected back almost exactly becoming quasiholes; the other side
of the vortex lets the quasiparticles to go through. Quasiholes are
reflected or transmitted in the opposite way. The vortex thus casts
a symmetric shadow for quasiparticles at one side and quasiholes at
the other, and, by measuring the flux of excitations, one detects
vortices and infer the vortex line density. A similar problem of
interaction of rotons with quantized vortices in $^4$He and
formation of shadows for $R^+$ and $R^-$ rotons was considered by
Samuels and Donnelly~\cite{Samuels}.

A related problem of Andreev reflection within the vortex core was
analyzed in Refs.~\cite{Kopnin,Stone} (see also the book of
Volovik~\cite{Volovik-book}). The analysis in cited works was
concerned with the bound states, whereas our concern is the
propagation of thermal excitations outside vortex cores.

In a recent paper \cite{Nugzar} we have solved analytically the
semi-classical equations of motion of ballistic quasiparticles in
the presence of a single stationary vortex. When extrapolated to a
disordered vortex tangle, our result is in agreement with simpler
order of magnitude estimates which have been used
\cite{Bradley-PRL2004,Fisher2008} to infer the vortex line density
in turbulence experiments. The aim of this article is to develop our
understanding of the interaction of ballistic quasi-particles and
vortices by considering more complex vortex configurations.

\section{Governing equations} \label{equations}

For the sake of simplicity, we consider the problem of motion of
quasiparticles in the $(x,\,y)$-plane in the presence of $N$
straight vortex lines aligned in the $z$ direction. The kinetic
energy of a thermal excitation of momentum $\pp$ measured with
respect to the Fermi energy $\eF$ is
\begin{equation}
\ep=\frac{p^2}{2m^*}-\eF,
\label{eq:ep}
\end{equation}
where $p=\vert \pp \vert$. Hereafter we use numerical values taken
at zero bar pressure~\cite{Wheatley} for the quantities which are
necessary to describe the motion of the excitation: the Fermi
velocity $\vF \approx 5.48 \times 10^3~\rm cm/s$, the Fermi momentum
$\pF=m^* \vF\approx 8.28 \times 10^{-20}~\rm g~ cm/s$, the Fermi
energy $\eF=\pF^2/(2m^*)\approx 2.27 \times 10^{-16}~ \rm erg$, and
the effective mass $m^* \approx 3.01 m =1.51 \times 10^{-23}~\rm g$,
where $m$ is the mass of the $^3$He atom.

Let $\Delta_0$ be the magnitude of the superfluid energy gap. Near
the vortex axis, at radial distances $r$ smaller than the
zero-temperature coherence length $\xi_0 = \hbar
v_F/\pi\Delta_0\approx0.85 \times 10^{-5}~\rm cm$, the energy gap
falls to zero and can be approximated by $\Delta(r) \approx
\Delta_0\tanh(r/\xi_0)$ \cite{Bardeen,TsunetoBook}. Since we are
mainly concerned with what happens to the excitation for $r \gg
\xi_0$, we neglect the spatial dependence of the energy gap and
assume the constant value, $\Delta_0=1.76k_B T_c \approx 2.43 \times
10^{-19}~\rm erg$, where $k_B$ is Boltzmann's constant and $T_c$ the
critical temperature.

The intersection of each vortex line with the $(x,\,y)$-plane is a
vortex point. Each vortex point moves with the flow field generated
by all other vortex points. The $i^{\rm th}$ vortex point, located
at the position ${\rr}_i(t)=x_i(t){\bf i}+y_i(t){\bf j}$ generates
the following velocity field at the point $\rr$:
\begin{equation}
{\vv}_i({\rr},\,t)=\frac{\kappa_i}{2\pi|{\rr}-{\rr}_i(t)|^2} [-{\bf
i}(y-y_i(t))+{\bf j}(x-x_i(t))], \label{eq:vel_vortex}
\end{equation}
where $\rr=x{\bf i}+y{\bf j}$, ${\bf i}$ and ${\bf j}$ are
respectively the unit vectors along the $x$ and $y$ axes, and the
circulation $\kappa_i$ of the $i^{\rm th}$ vortex is $\kappa_i=\pm
\kappa$; the $+$ and $-$ signs denote respectively a vortex
(anticlockwise rotation in the $(x,\,y)$-plane) and an antivortex
(clockwise rotation). The quantity
\begin{equation}
\kappa=\frac{h}{2m}= \frac{\pi\hbar}{m}=  0.662\times
10^{-3}~\textrm{cm}^2/\textrm{s} \label{eq:kappa}
\end{equation}
is the quantum of circulation in $^3$He-B. The velocity field at the
point $\rr $ created by the $N$ vortices is
\begin{equation}
\vv_s(\rr,\,t)=\sum_{i=1}^{i=N} \vv_i(\rr,\,t), \label{eq:vel_field}
\end{equation}
thus the velocity of the $i^{\rm th}$ vortex point $\rr_i$ is
\begin{equation}
\frac{d\rr_i(t)}{dt}=\sum_{j=1,j\neq i}^{i=N} \vv_j(\rr_i).
\label{eq:vel_ipoint}
\end{equation}

In the presence of vortices the energy of the thermal excitation becomes
\begin{equation}
E=\sqrt{\ep^2+\Delta_0^2}+\pp \cdot \vv_s(\rr,\,t). \label{eq:E}
\end{equation}
In writing Eq.~(\ref{eq:E}), the spatial variation of the order
parameter is not taken into account for the sake of simplicity. We
also assume that the interaction term $\pp \cdot \vv_s$ varies on a
spatial scale which is larger than $\xi_0$, and that the excitation
can be considered a compact object of momentum $\pp=\pp(t)$,
position $\rr=\rr(t)$, and energy $E=E(\pp,\,\rr,\,t)$. This gives
us the opportunity to use the method developed in
Ref.~\cite{Leggett}, and consider Eq.~(\ref{eq:E}) as an effective
Hamiltonian, for which the equations of motion are
\begin{equation}
\frac{d\rr}{dt}=\frac{\partial E(\pp,\rr)}{\partial \pp}
=\frac{\ep}{\sqrt{\ep^2+\Delta_0^2}} \frac{\pp}{m^*} + \vv_s,
\label{eq:hamilton1}
\end{equation}
\begin{equation}
\frac{d\pp}{dt}=-\frac{\partial E(\pp,\rr)}{\partial \rr}
=-\frac{\partial}{\partial \rr}(\pp \cdot \vv_s).
\label{eq:hamilton2}
\end{equation}
Eq.~(\ref{eq:hamilton1}) represents the group velocity of the
excitation in the velocity field of the vortices. Excitations such
that $\ep>0$ are called quasiparticles, and excitations such that
$\ep<0$ are called quasiholes. The right-hand-side of
Eq.~(\ref{eq:hamilton2}) is thus the force acting on the excitation.

Before solving numerically Eqs.~(\ref{eq:hamilton1}) and
(\ref{eq:hamilton2}) it is convenient to rewrite them in
dimensionless form. We introduce the following dimensionless
variables:
\begin{equation}
H=\frac{E}{\Delta_0},
\label{eq:dimless-E}
\end{equation}
\begin{equation}
{\bf{\Pi}}=\frac{{\pp}}{\pF},
\label{eq:dimless-P}
\end{equation}
\begin{equation}
{\bf V}_s=\frac{\xi_0}{\kappa}\vv_s,
\label{eq:dimless-Vs}
\end{equation}
\begin{equation}
{\bf R}=(X,\,Y)=\biggl(\frac{x}{\xi_0},\,\frac{y}{\xi_0}\biggr)=
\frac{\rr}{\xi_0}, \label{eq:dimless-R}
\end{equation}
\begin{equation}
\tau=t/t_0,
\label{eq:dimless-t}
\end{equation}
where $t_0=\xi_0 \pF/\Delta_0=2.9 \times 10^{-6}~{\rm s}$. The
Hamiltonian, Eq.~(\ref{eq:E}) and the equations of motion,
Eqs.~(\ref{eq:hamilton1}) and (\ref{eq:hamilton2}) then become:
\begin{equation}
H({\bf\Pi},{\bf R},\tau)= \lambda\sqrt{({\Pi}^2-1)^2+\lambda^{-2}}+
\frac{m^*}{m}\pi^2{\bf\Pi}\cdot{\bf V}_s({\bf
R},\tau),\label{eq:dimhamilt}
\end{equation}
and
\begin{equation}
\frac{dX}{d\tau}=
\lambda\frac{2(\Pi^2-1)}{\sqrt{(\Pi^2-1)^2+\lambda^{-2}}}\Pi_x+
\frac{m^*}{m}\pi^2 V_{sx}({\bf R},\,\tau),
\label{eq:dimhamilton1}
\end{equation}
\begin{equation}
\frac{dY}{d\tau}=
\lambda\frac{2(\Pi^2-1)}{\sqrt{(\Pi^2-1)^2+\lambda^{-2}}}\Pi_y+
\frac{m^*}{m}\pi^2 V_{sy}({\bf R},\,\tau),
\label{eq:dimhamilton2}
\end{equation}
\begin{equation}
\frac{d\Pi_x}{d\tau}=
-\frac{m^*}{m}\pi^2
\biggl(\Pi_x\frac{dV_{sx}}{dX}+
\Pi_y\frac{dV_{sy}}{dX}\biggr),
\label{eq:dimhamilton3}
\end{equation}
\begin{equation}
\frac{d\Pi_y}{d\tau}=
-\frac{m^*}{m}\pi^2
\biggl(\Pi_x\frac{dV_{sx}}{dY}+
\Pi_y\frac{dV_{sy}}{dY}\biggr),
\label{eq:dimhamilton4}
\end{equation}
where the dimensionless parameter $\lambda$ is
\begin{equation}
\lambda=\frac{\eF}{\Delta_0}.
\label{eq:lambda}
\end{equation}
In our numerical calculations we shall assume the value
$\lambda=10^3$. Finally, the dimensionless superfluid velocity is
\begin{equation}
{\bf V}_s({\bf R},\,\tau)=\sum_{i=1}^{i=N}{\bf V}_i({\bf R},\,\tau)=
\sum_{i=1}^{i=N}\frac{\Gamma_i}{2\pi|{\bf R}-{\bf R}_i(\tau)|^2}
[-{\bf i}(Y-Y_i(\tau))+
{\bf j}(X-X_i(\tau))],
\label{eq:dimvelfield}
\end{equation}
where $\Gamma_i=1$ for vortices, $\Gamma_i=-1$ for antivortices, and
\begin{equation}
\frac{d{\bf R}_i(\tau)}{d\tau}=\sum_{j=1,j\neq i}^{j=N}{\bf V}_j({\bf R}_i).
\end{equation}

\section{Single vortex} \label{single}

The numerical solution of
Eqs.~(\ref{eq:dimhamilton1})-(\ref{eq:dimhamilton4}) which govern the trajectory of
quasiparticles and vortices requires special care due to the
absence of dissipation mechanisms. Our preliminary investigations revealed
that the most commonly used
differential equation solvers, such as for
example the Runge-Kutta fourth order method, are not satisfactory, even
using a very small time step;
in the case of a single vortex, these solvers failed to
conserve the total energy and the total angular momentum of the
quasiparticle by large
amounts (10\% or more). In the
case of more complex, time dependent vortex configurations, energy and
momentum of quasiparticles would not be conserved, but clearly we
could not trust our results if the basic conservation laws were
not satisfied in the simplest case of a single vortex.

Ideally, to build the conservation law into the numerical scheme,
the numerical method must be symplectic and conserve  phase-space
volume \cite{Sanz-Serna}. Unfortunately the known symplectic
algorithms are geared to problems (mainly in the context of gravity)
in which the Hamiltonian has the additive form $H=T(\pp)+V(\qq)$,
where $\pp$ and $\qq$ are the generalized momenta and positions, $T$
is the kinetic energy, and $V$ the potential energy, whereas in our
problem the variables $\pp$ and $\qq$ appear in nonlinear
combinations. The second difficulty is the stiffness of our
equations of motion, as very rapid time-scales appear at the Andreev
turning points. After some experimenting, we have found that we can
solve the governing equations with satisfactory accuracy using the
Matlab code {\it ode15s}, which is a quasi-constant step size
implementation of the numerical differentiation formulas (NDF)
particularly efficient for solving stiff problems (for detailed
description of the {\it ode15s} Matlab solver and corresponding
software see Ref.~\cite{ode15s}). When solving
Eqs.~(\ref{eq:dimhamilton1})-(\ref{eq:dimhamilton4}), error
tolerances were lowered until the particle trajectory had
sufficiently converged, in particular at reflections.

To test our numerical method we determine the trajectories of
excitations in the presence of a single
vortex located at the origin, and compare the results with
previous analytical results \cite{Nugzar}.
The velocity field of the vortex is simply
\begin{equation}
{\bf V}_s({\bf R})=
\frac{1}{2\pi R^2}
(-{\bf i}Y+
{\bf j}X). \label{eq:singvorvelfield}
\end{equation}
Since the vortex does not move, this velocity field
is time-independent, and the governing equations
(\ref{eq:dimhamilton1})-(\ref{eq:dimhamilton4})
have two integrals of motion: the first is the energy, defined
by the Hamiltonian, Eq.~(\ref{eq:dimhamilt}); the second
is the $z$-component of the angular momentum, which is
\begin{equation}
J=\Pi_{y}X-\Pi_{x}Y.
\end{equation}

\begin{figure}
%\centering \epsfig{figure=shadow_2_16_colour.eps, height=3.5in,angle=0}
\centering \epsfig{figure=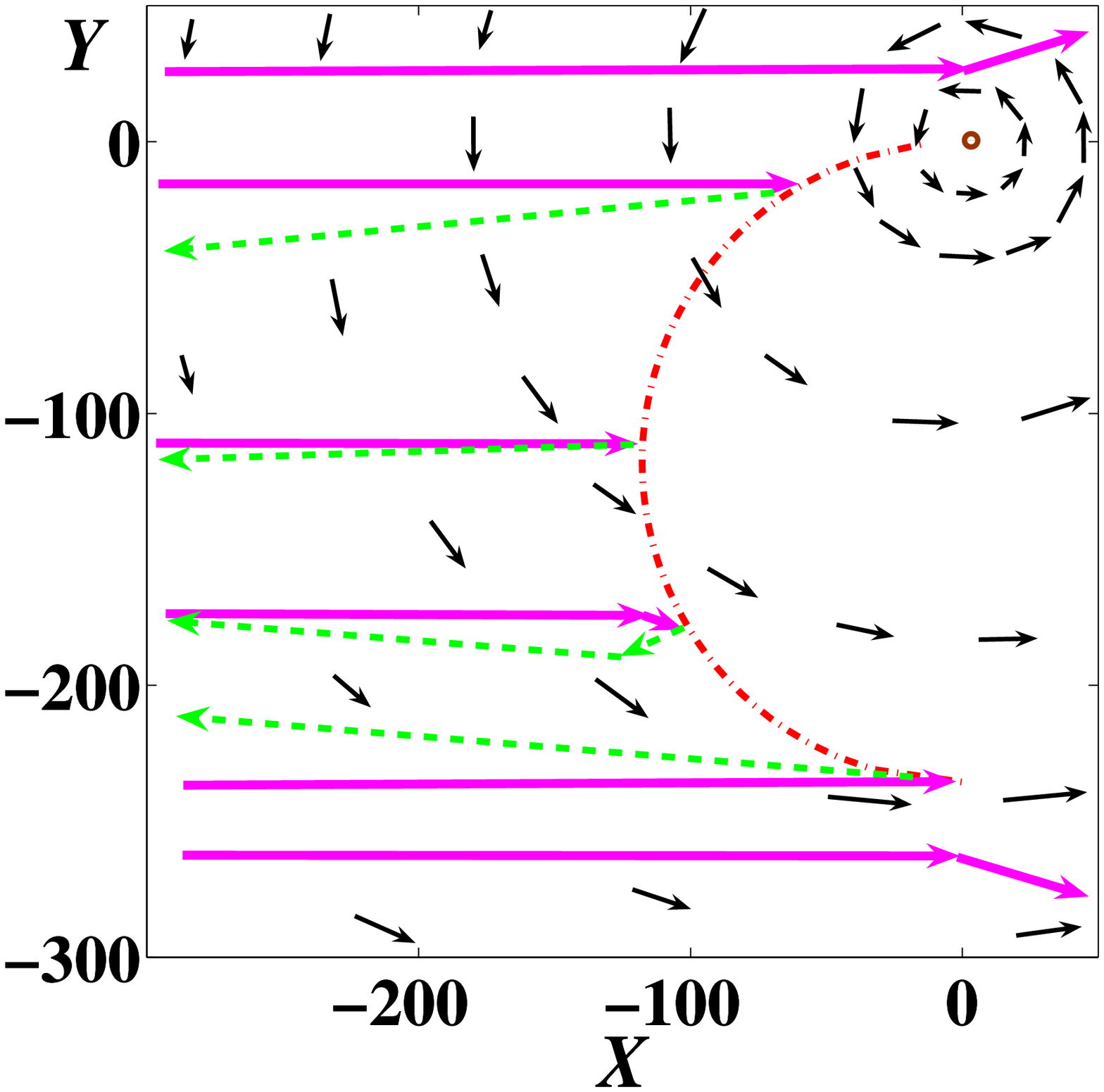, height=3.5in,angle=0}
\caption{(Color online). Trajectories of excitations with initial
momentum ${\bf \Pi}_0=(1.0001,0)$ and initial position $(X_0,Y_0)$
with $X_0=-10^4$ for different values of $Y_0$ in the presence of a
single (positive) vortex at the origin (marked by the dot). The
(anticlockwise) superfluid velocity field of the vortex is indicated
by arrows. Quasiparticles trajectories are solid (purple) lines,
quasiholes trajectories are dashed (green) lines. The dashed-dotted
(red) curve denotes the locus of Andreev reflection.} \label{fig:1}
\end{figure}

\begin{figure}[t]
\begin{tabular}[b]{cc}
\includegraphics[height=0.37\linewidth]{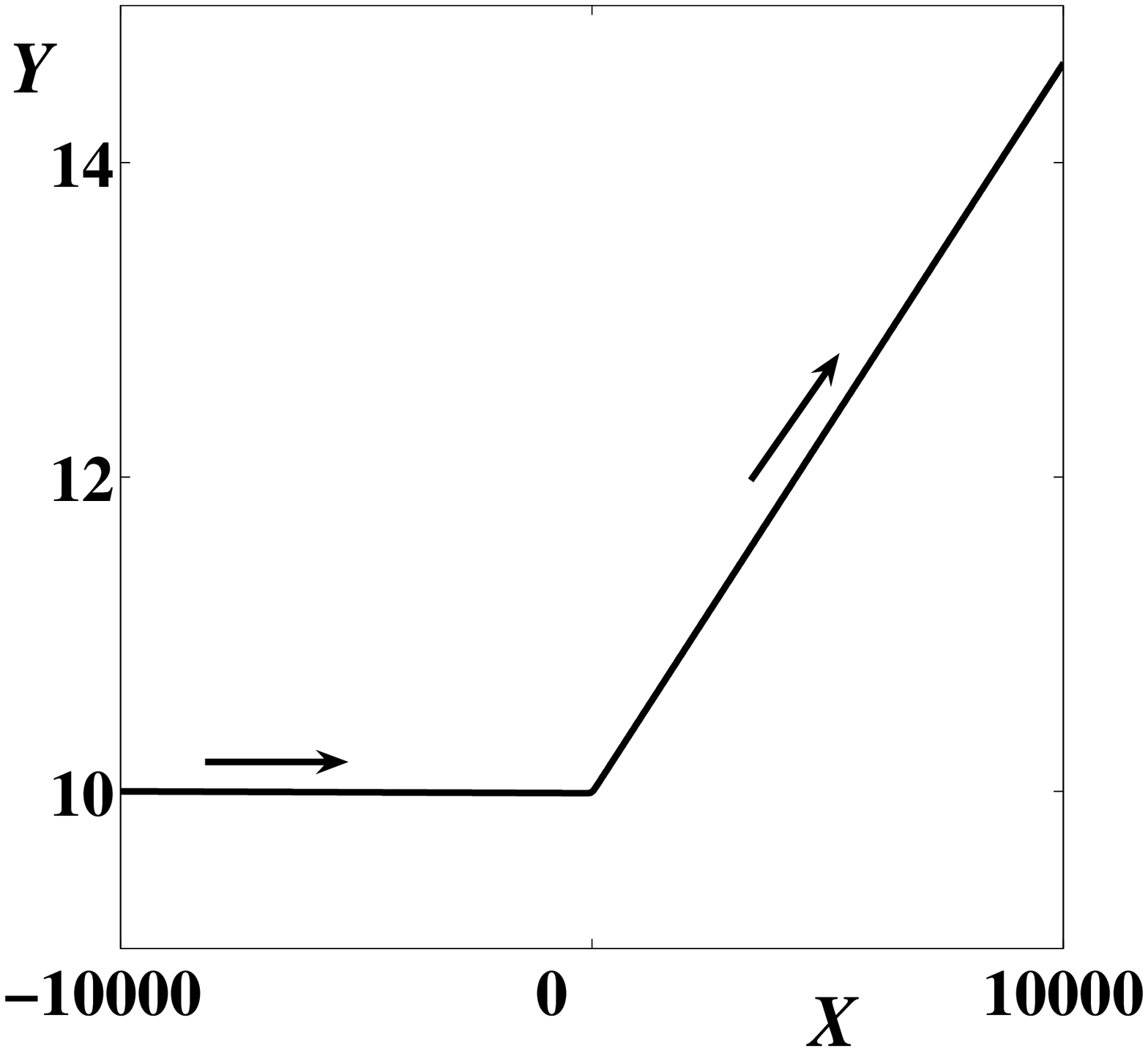}&
\includegraphics[height=0.37\linewidth]{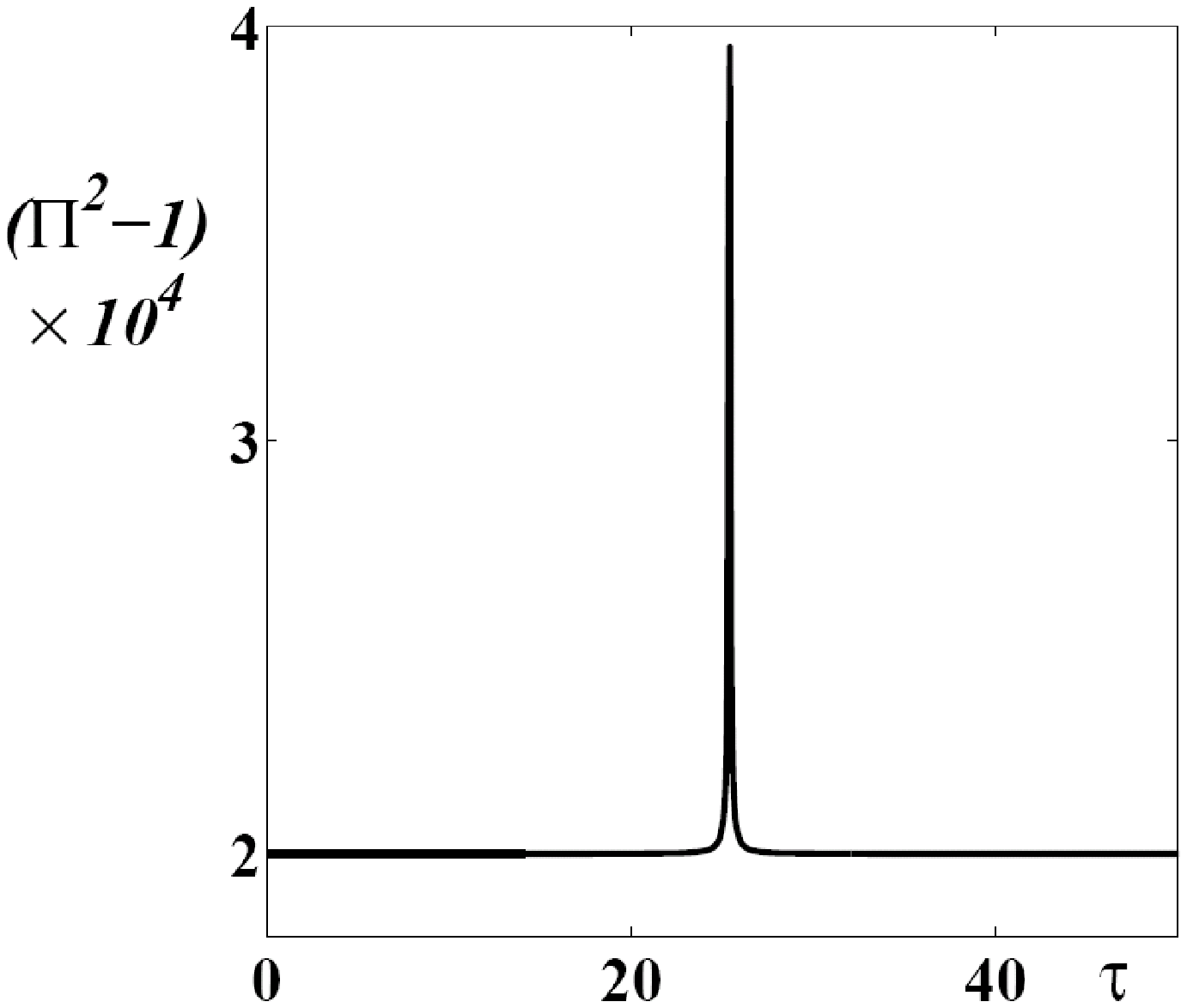}\\
\end{tabular}
\caption{Left: trajectory of the quasiparticle with initial momentum
${\bf \Pi}_0=(1.0001,\,0)$ and position $(X_0,\,Y_0)=(-10^4,\,10)$
in the presence of a single anticlockwise vortex at the origin. The
arrows indicate the direction of motion. Right: plot of $\Pi^2-1$ vs
time $\tau$ corresponding to the left part of the figure; note that
the quasiparticle remains a quasiparticle.} \label{fig:2}
\end{figure}

\begin{figure}
%\centering \epsfig{figure=integrals_2_3.eps,height=3.5in,angle=0}
\centering \epsfig{figure=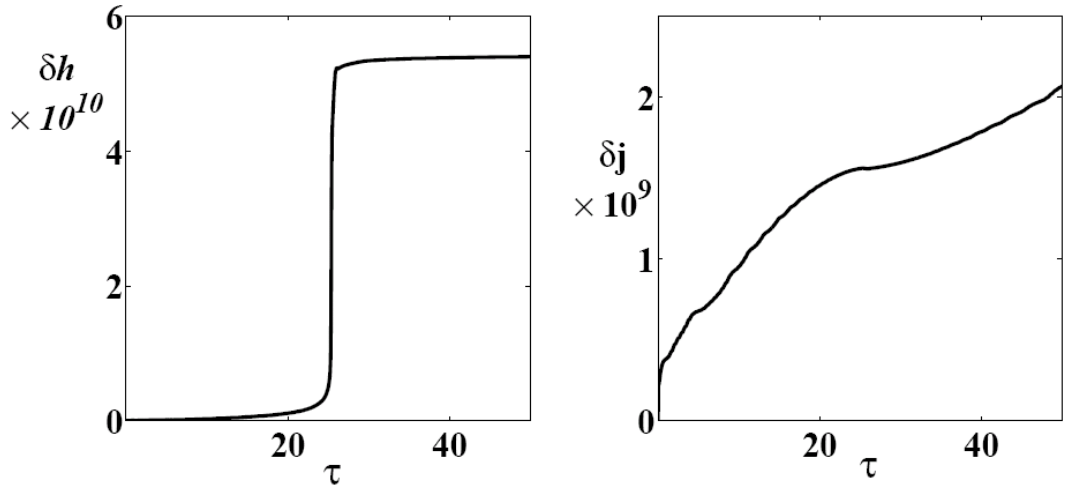,height=3in,angle=0}
\caption{Plot of relative energy variation $\delta h=(H-H_0)/H_0$ vs
time $\tau$ (left) and relative angular momentum variation $\delta
j=(J-J_0)/J_0$ (right) corresponding to Fig.~\ref{fig:2}.}
\label{fig:3}
\end{figure}

The initial conditions at $\tau=0$ for our calculations are the
following. The initial momentum is ${\bf \Pi}_0=(1.0001,~0)$
and corresponds to a quasiparticle moving in the $x$ direction with
energy $E=\Delta_0+k_BT$ for temperature $T\approx 0.1 T_c$.
The initial position is $(X_0,Y_0)$ with $X_0=-10^4$, far away from
the vortex. We study the trajectory of the quasiparticle
as a function of $Y_0$, which plays the role of impact parameter.
Fig.~\ref{fig:1} shows results for some typical values of $Y_0$.
We distinguish three cases:
\bigskip

\noindent
{\bf Case 1:}
For $Y_0\ge 0$ we have no reflection,
in agreement with previous work~\cite{Nugzar}.
For example, Fig.~\ref{fig:2} (left) shows the
quasiparticle's trajectory for $Y_0=10$;
Fig.~\ref{fig:2} (right) shows that
$\Pi^2-1$ remains positive at all times $\tau$, which means
that the quasiparticle retains its nature of quasiparticle.
Fig.~\ref{fig:3} confirms that our numerical method conserves
energy and angular momentum very well.
The left hand side
of Fig.~\ref{fig:3} shows that the relative error in the energy,
$\delta h(\tau)=(H(\tau)-H_0)/H_0$, where $H_0$ is the initial energy,
is less than $6 \times 10^{-10}$;
the right hand side of Fig.~\ref{fig:3} shows that the relative error
in computing the angular momentum,
$\delta j(\tau)=(J(\tau)-J_0)/J_0$, where $J_0$ is the
initial angular momentum, is less than $2.5 \times 10^{-9}$.

\begin{figure}[t]
\begin{tabular}[b]{cc}
\includegraphics[height=0.38\linewidth]{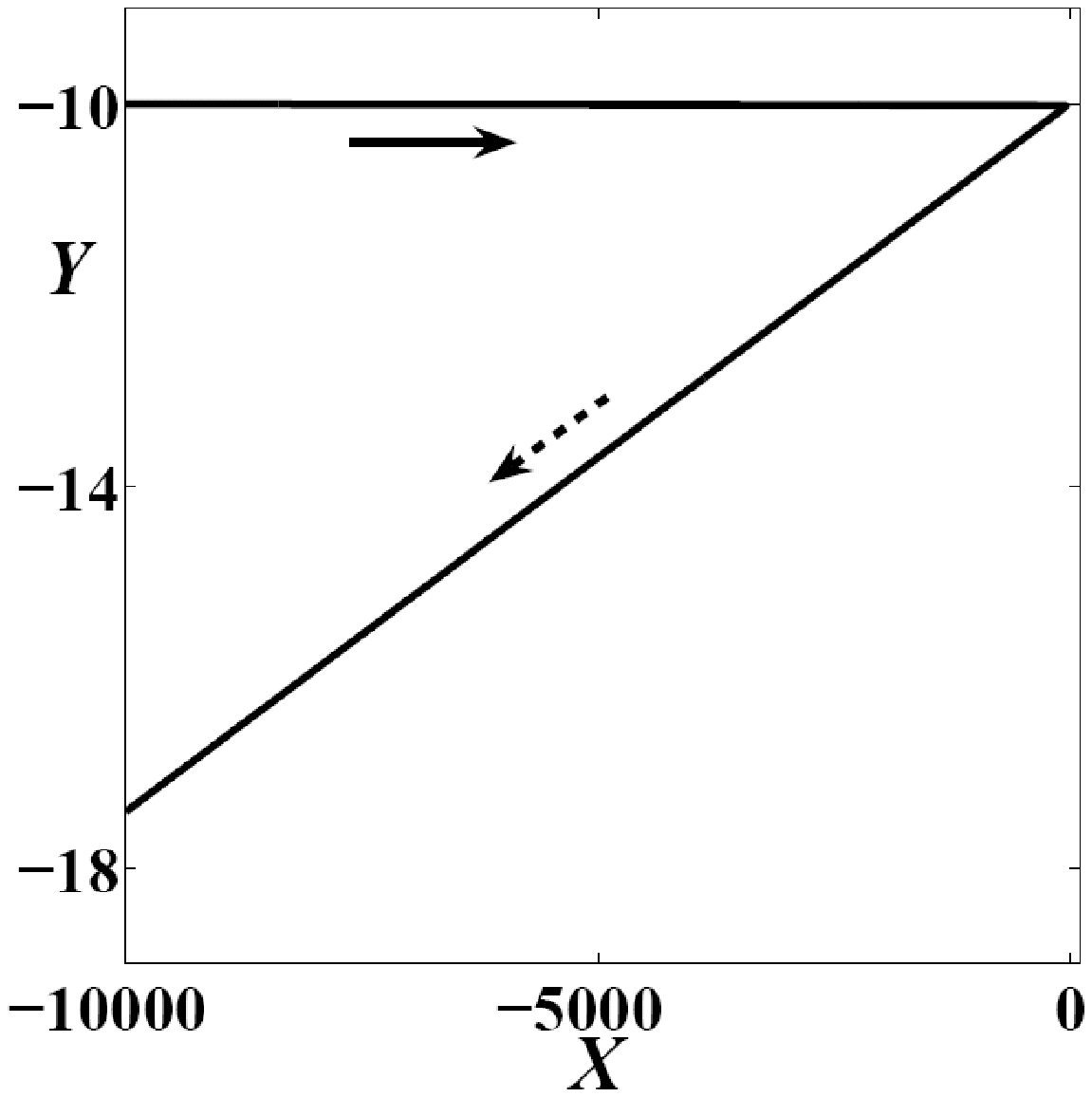}&
\includegraphics[height=0.385\linewidth]{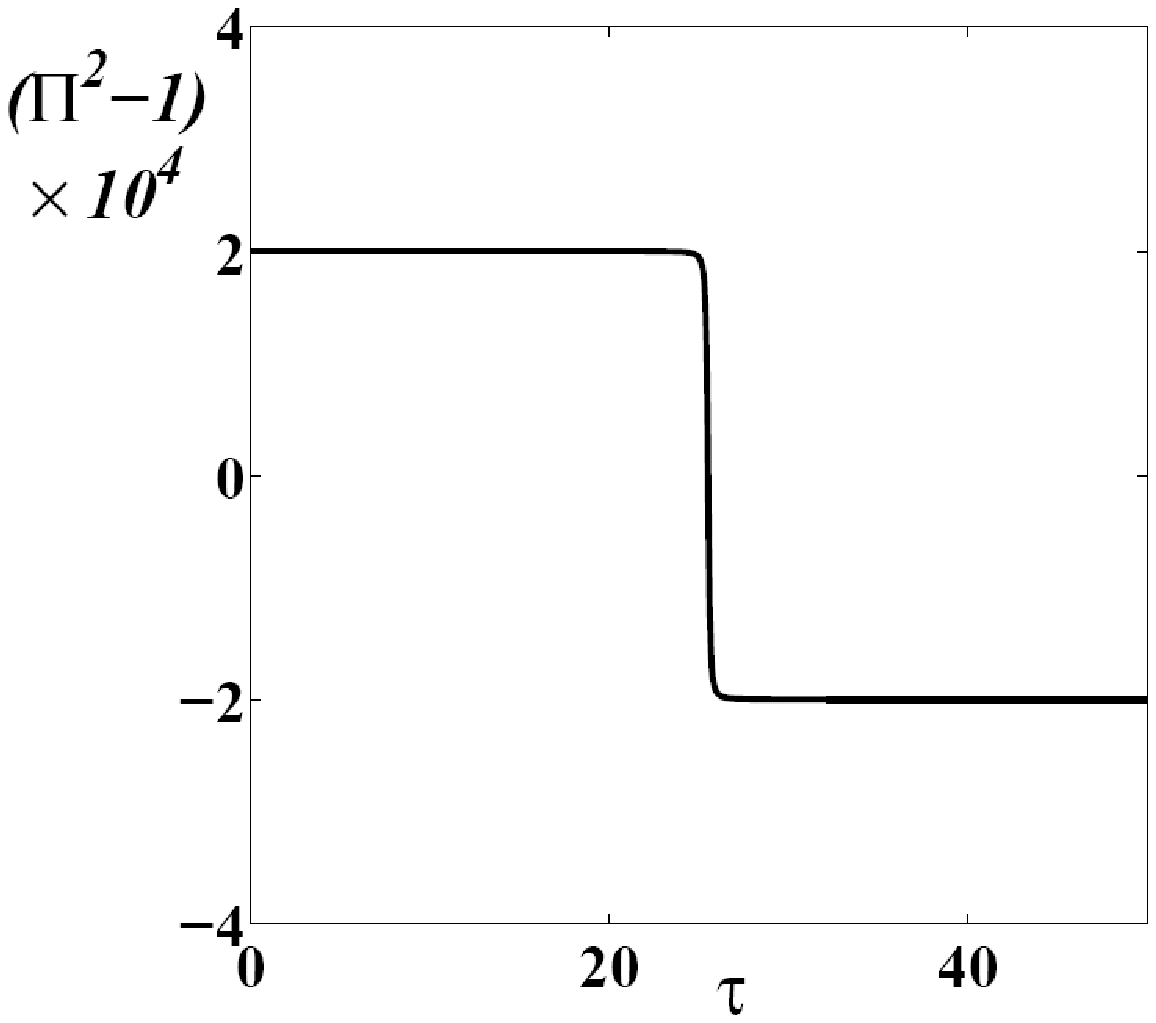}\\
\end{tabular}
\caption{Left: trajectory of quasiparticle with initial momentum
${\bf \Pi}_0=(1.0001,\,0)$ and position $(X_0,\,Y_0)=(-10^4,-10)$ in
the presence of a single anticlockwise vortex at the origin. The
arrows indicate the direction of motion. Right: plot of $\Pi^2-1$ vs
time $\tau$ corresponding to the left part of the figure; note that
the quasiparticle becomes a quasihole.} \label{fig:4}
\end{figure}

\begin{figure}[t]
\begin{tabular}[b]{cc}
\includegraphics[height=0.38\linewidth]{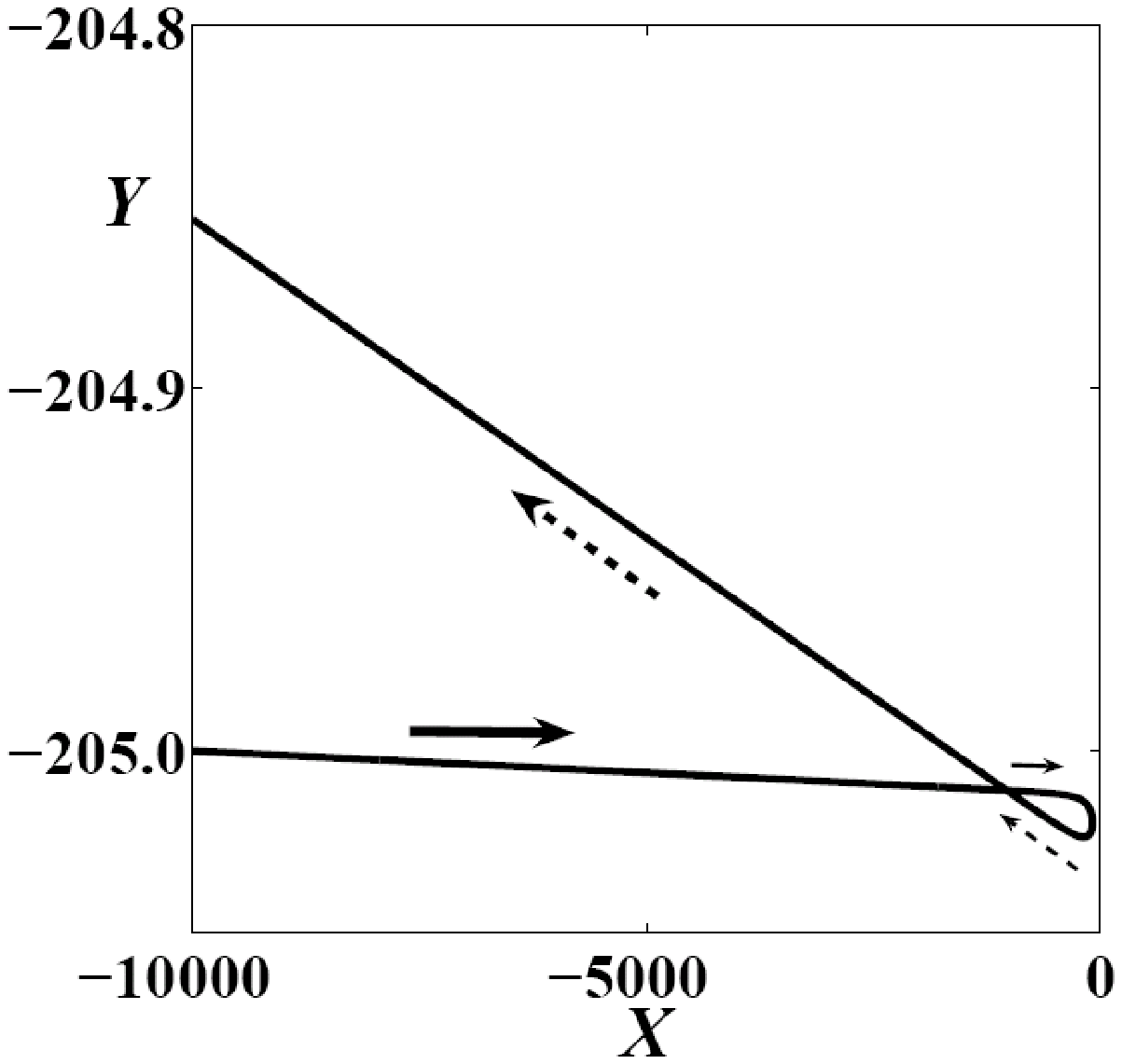}&
\includegraphics[height=0.385\linewidth]{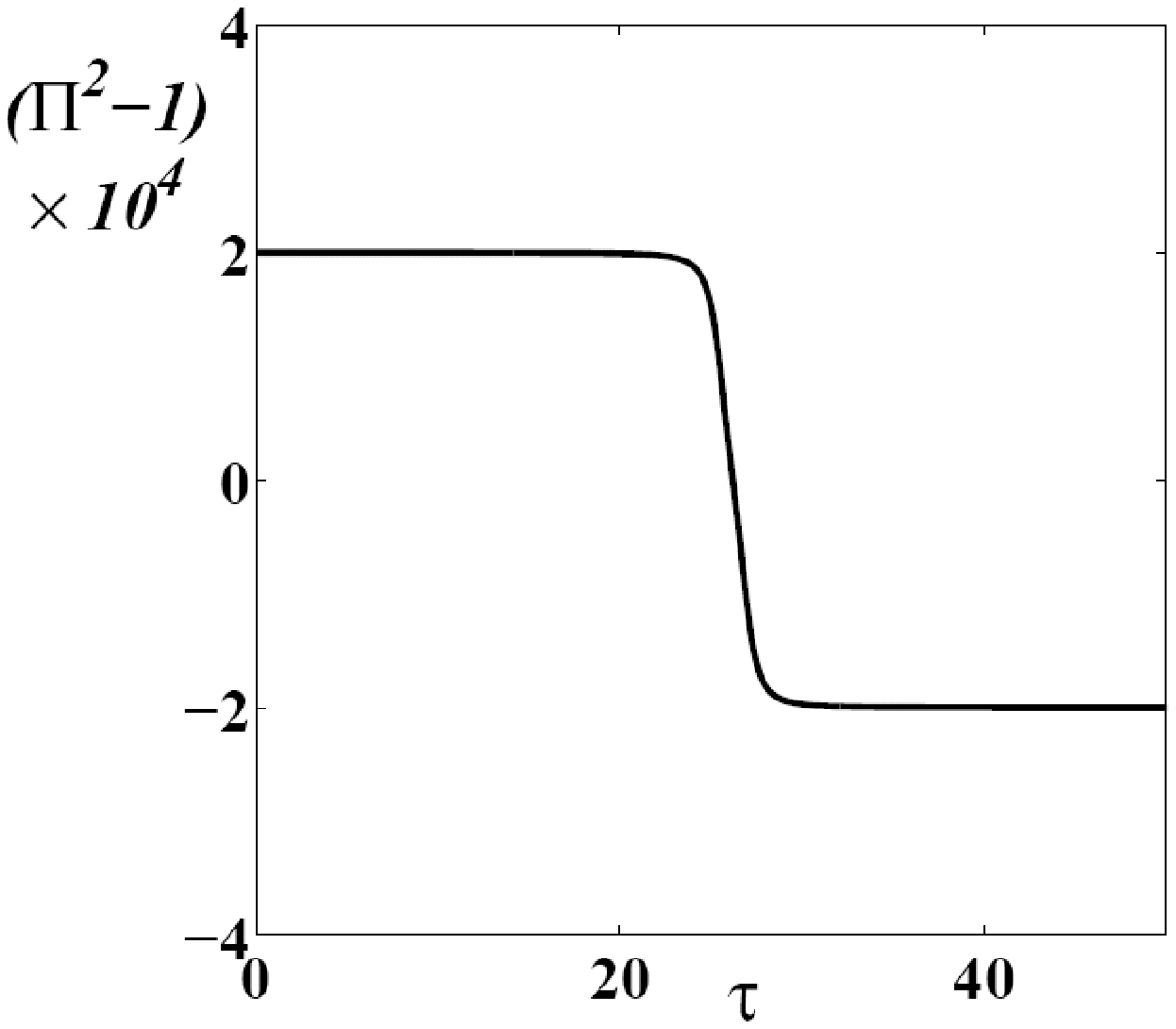}\\
\end{tabular}
\caption{Left: trajectory of the quasiparticle with initial momentum
${\bf \Pi}_0=(1.0001,\,0)$ and position $(X_0,\,Y_0)=(-10^4,\,-205)$
in the presence of a single anticlockwise vortex at the origin. The
arrows indicate the direction of motion. Right: plot of $\Pi^2-1$ vs
time $\tau$ corresponding to the left part of the figure; note that
the quasiparticle becomes a quasihole.} \label{fig:5}
\end{figure}

\bigskip
\noindent
{\bf Case 2:}
If $Y_0<0$ but $\vert Y_0 \vert$ is not too large, the
incident
quasiparticle is Andreev reflected, as shown for
example in Fig.~\ref{fig:4} (left) for $Y_0=-10$.
Fig.~\ref{fig:4} (right)
shows that $\Pi^2-1$ changes sign, thus confirming that, upon
reflection, the quasiparticle  becomes a quasihole. In this
calculation, the numerical errors in conserving
energy and angular momentum are $\delta h < 8 \times 10^{-10}$
and $\delta j < 2 \times 10^{-9}$ respectively. Fig.~\ref{fig:5}
shows another Andreev reflection, this time for $Y_0=-205$.

In our previous paper \cite{Nugzar} we determined the distance from
the vortex at which, if $Y_0<0$, the incident quasiparticle is
Andreev reflected; the dashed-dotted (red) curve in Fig.~\ref{fig:1}
marks this location. It is apparent that there is a maximum value of
$\vert Y_0 \vert$ past which a quasiparticle with energy
$\epsilon_p$ is not Andreev reflected; this value (in our
dimensionless units) is approximately equal to $S_0=3 \pi
(\Delta_0/\epsilon_p)^2\approx 269$ where we used $\epsilon_p
=\epsilon_F (\Pi^2-1) \approx 0.0002 \epsilon_F$ for $\Pi=1.0001$.
We call $S_0=269$  the (dimensionless) {\it Andreev shadow} of a
single  vortex to quasiparticles of that particular (dimensionless)
momentum $\Pi$.
\bigskip

\noindent
{\bf Case 3:}
Finally, if $0>-S>Y_0$, the quasiparticle's trajectory is deflected
by the vortex but remains a quasiparticle.

\section{Vortex-Vortex Pair} \label{pair}

The velocity field of two vortices is time-dependent,
thus the Hamiltonian of the thermal excitation has no
integrals of motion. Unlike the previous case
of a single vortex, $H$ and $J$ are not conserved. The only
quantity in the problem which is constant is the
the distance between the vortices. There are two cases to
consider: two vortices of the same circulation (vortex-vortex pair),
and two vortices of the opposite circulations (vortex-antivortex pair).
This section is concerned with the former.
\bigskip

Two vortices of the same circulation at distance $d$ from
each other rotate around a point halfway
between them with velocity
\begin{equation}
v_{rot}=\frac{\kappa}{2\pi d}.
\label{eq:rot_vel}
\end{equation}
In dimensionless variables we have
\begin{equation}
V_{rot}=\frac{1}{2\pi D},~~~~~D=\frac{d}{\xi_0}.
\label{eq:rot_vel_dim}
\end{equation}

Far from the vortices, the velocity of the quasiparticle can be
estimated from Eqs.~(\ref{eq:dimhamilton1})
and (\ref{eq:dimhamilton2}):
\begin{equation}
\frac{d{\bf R}}{d \tau}\approx 2\lambda^2(\Pi^2-1){\bf \Pi}.
\end{equation}

To get a more clear picture of the problem, it is useful to make the
following simple estimates. Away from the vortices, the velocity of
the quasiparticle is approximately $400$. If $D=10$, the velocity of
the vortices is approximately $0.016$. For a typical timescale of
approximately $25$ to $30$ time units, the vortices travel
approximately  the distance $0.4$ to $0.5$, which means that they
rotate about the center of rotation by an angle $\delta \phi \approx
2.3^o$ to $3^o$. If $D$ is larger than $10$ the vortices move slower
and $\delta \phi$ is even smaller. We conclude that, in the first
approximation, the vortex system is static to quasiparticles with
the momentum ${\bf\Pi}_0=(1.0001,\,0)$. However, if
${\bf\Pi}_0=(1.00005,\,0)$, in the corresponding time scale the
vortices move by a more substantial angle, $\delta \phi \approx
4.5^o$ to $6^o$. If $D=10$, a significant displacement can be
observed for quasiparticles with ${\bf\Pi}_0=(1.00002,\,0)$ and the
vortex configuration cannot be considered static.

\begin{figure}[ht]
\centering
\includegraphics[angle=0,height=2.6in]{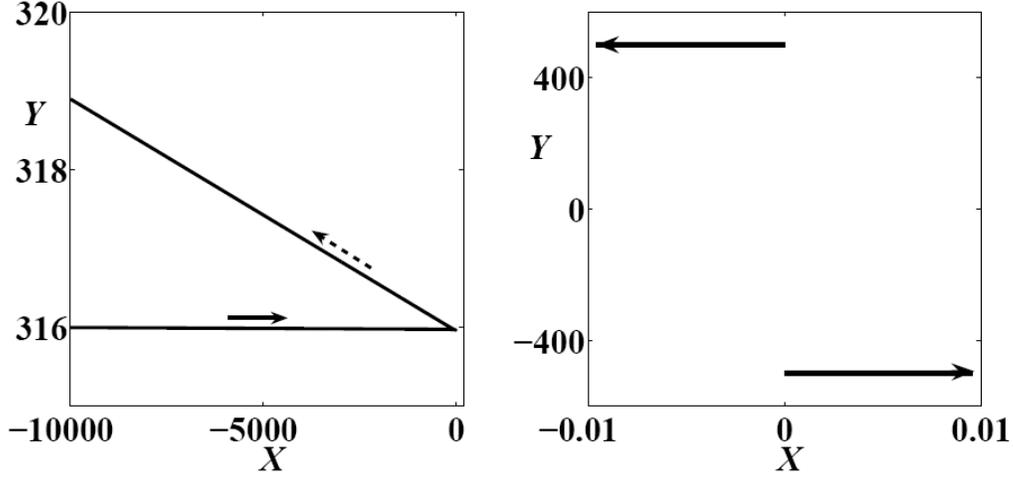}
\caption{Left: trajectory of the quasiparticle with initial momentum
${\bf \Pi}_0=(1.0001,\,0)$ and position $(X_0,\,Y_0)=(-10^4,\,316)$
in the presence of a vortex-vortex pair; the first vortex is at
$(Q_{x1},\,Q_{y1})=(0,\,-500)$, and the second vortex is at
$(Q_{x2},\,Q_{y2})=(0,\,500)$. The solid arrow indicates the
incident quasiparticle, the dashed arrow the reflected quasihole.
Right: Directions of motion of the two vortices.} \label{fig:6}
\end{figure}

\begin{figure}[t]
\begin{tabular}[b]{cc}
\includegraphics[height=0.37\linewidth]{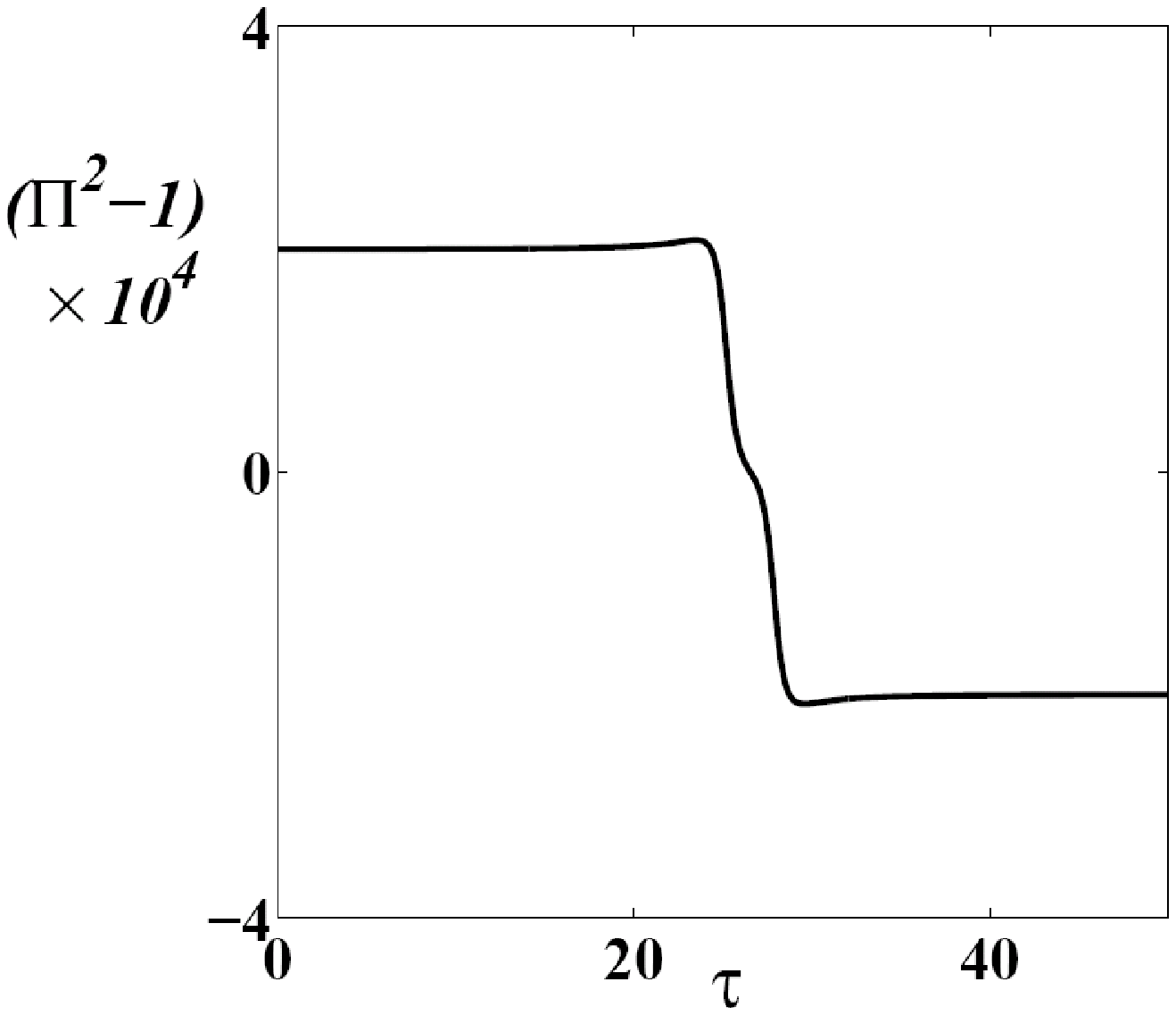}&
\includegraphics[height=0.37\linewidth]{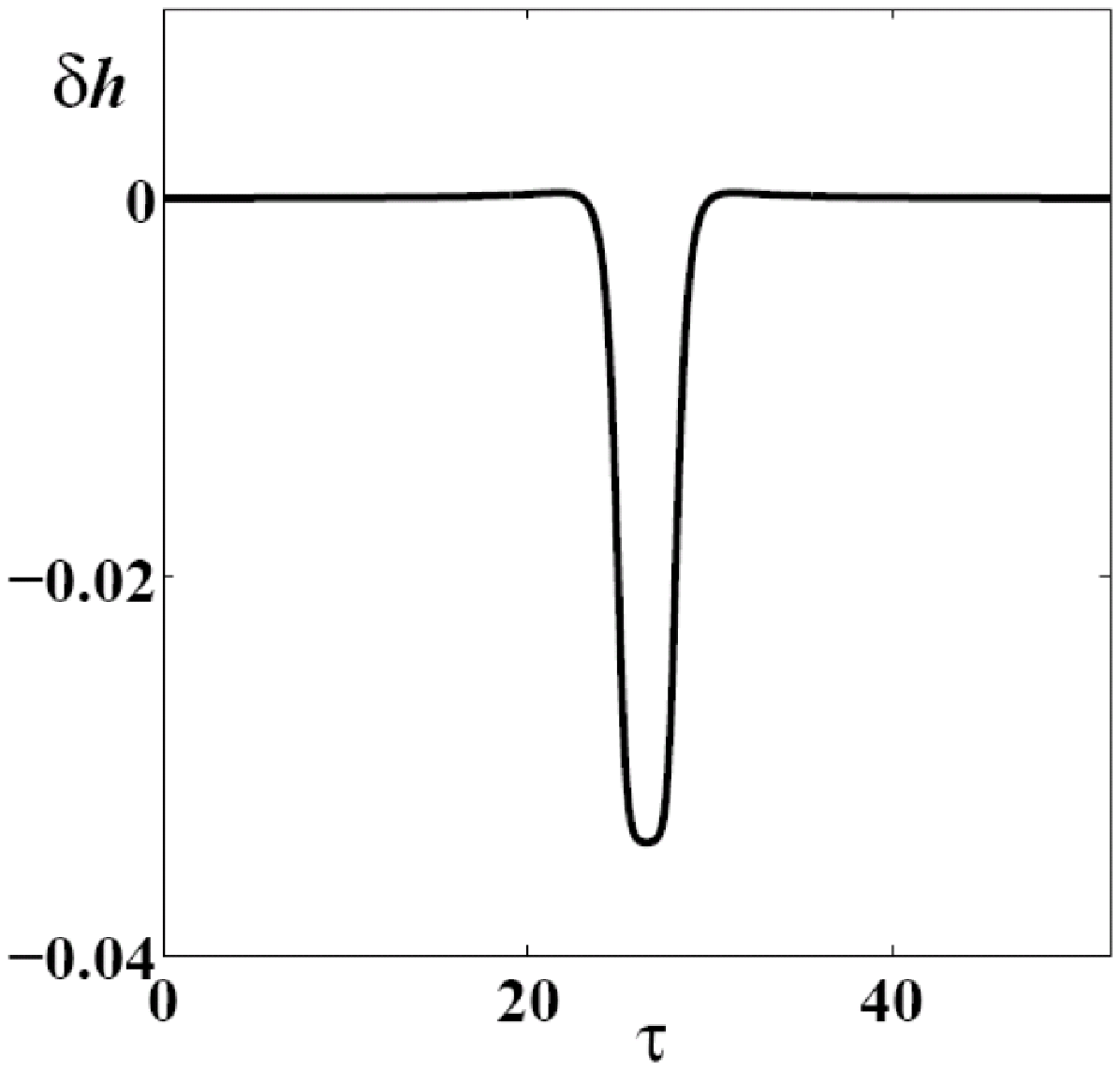}\\
\end{tabular}
\caption{Left: plot of $\Pi^2-1$ vs time $\tau$ corresponding to the
left part of Fig.~\ref{fig:6}; note that the quasiparticle becomes a
quasihole. Right: plot of relative energy difference $\delta
h=(H(\tau)-H_0)/H_0$ vs time $\tau$.} \label{fig:7}
\end{figure}

We proceed with our calculations and consider two vortices of the
same circulation at distance $D=1000$ from each other, initially
located at positions $(Q_{x1},\,Q_{y1})=(0,\,-500)$ and
$(Q_{x2},\,Q_{y2})=(0,\, 500)$. We integrate the equations of motion
of quasiparticles with ${\bf\Pi}_0=(1.0001,\,0)$, keeping
$X_0=-10^4$ fixed and varying $Y_0$. In this configuration, the
relative angle $\theta$ between the direction of the incoming
quasiparticle (the $x$ axis) and the line which joins the two
vortices (the $y$ axis) is $\theta=90^o$. Fig.~\ref{fig:6}
illustrates the trajectories of the quasiparticle for $Y_0=316$ as
well as the path of the vortices; Fig.~\ref{fig:7} (left) shows that
the incident quasiparticle is reflected as a quasihole; since the
$\pp \cdot {\bf v}_s$ term in the Hamiltonian is time-dependent, the
energy is not constant during the evolution, as confirmed in
Fig.~\ref{fig:7} (right). We find that the Andreev shadow of the
first vortex of the pair is $S_1=290$, slightly more than $S_0$ (the
Andreev shadow of an isolated stationary vortex), and that the
shadow of the second vortex of the pair is $S_2=184$, which is
slightly less than $S_0$. Note that in this case $S_1+S_2 \approx 2
S_0$.

If the relative angle $\theta$
between the direction of the incoming quasiparticle
and the direction between the vortices is reduced from $\theta=90^o$
to $\theta=45^o$, the
Andreev shadow of the first vortex decreases from $S_1=290$ to
$S_1=272$, but the Andreev shadow of the second vortex increases
from $S_2=184$ to $S_2=205$, so that the total shadow $S_1+S_2$
of the vortex configuration
remains approximately equal to twice the shadow $S_0$ of a single
isolated vortex.
For example, if the vortices are located
at $(Q_{x1},\,Q_{y1})=(-353,\, -353)$ and
$(Q_{x2},\,Q_{y2})=(353,\, 353)$, the angle is $\theta=45^o$,
$S_1=272$, $S_2=205$, and $S_1+S_2=477$. If $\theta$ is further
reduced from $\theta=45^o$ to $14.1^o$, the total shadow of both
vortices increases and reaches its maximum size:
$S_1=263$, $S_2=242$, and $S_1+S_2=505$. If $\theta>14.1^o$ the
two shadows merge, and the two vortices screen each other.
If $\theta=0^o$ then $S_1+S_2=251$ only.

Qualitatively, the same behaviour is observed if the distance between
the vortices is even smaller, e.g. $D=100$. If the angle
between the direction of motion of the excitation and the
direction of the line through the vortices is $\theta=90^o$,
the shadow of the first vortex at $(Q_{x1},\,Q_{y1})=(0,\, 50)$ is
only $S_1=44.7$ (significantly smaller than $S_0$)
and the shadow of the second vortex at $(Q_{x2},\,Q_{y2})=(0,\,-50)$ is
$S_2=443.3$ (significantly larger than $S_0$), and the total
shadow is $S_1+S_2=488$. If $\theta$ is reduced the two shadows merge
when $\theta \approx 75^o$; at this angle $S_1+S_2=526$.
At $\theta=45^o$, the total shadow is $S_1+S_2=510$, and at $\theta=0^o$
$S_1+S_2=469$.

What happens if $D$ is further reduced? Consider for simplicity
the angle $\theta=90^o$: for large $D$ the two vortices have independent
shadows (one reduced in size, the other increased in size).
By decreasing the distance between the vortices the
shadows approach each other; for distances $D\lesssim 16$ they merge and
the vortex configuration has a single shadow independent of $\theta$,
as if it were a single vortex of strength approximately equal to
$2 \kappa$.

\section{Vortex-antivortex pair} \label{antipair}

\begin{figure}
\centering
\includegraphics[angle=0,height=2.5in]{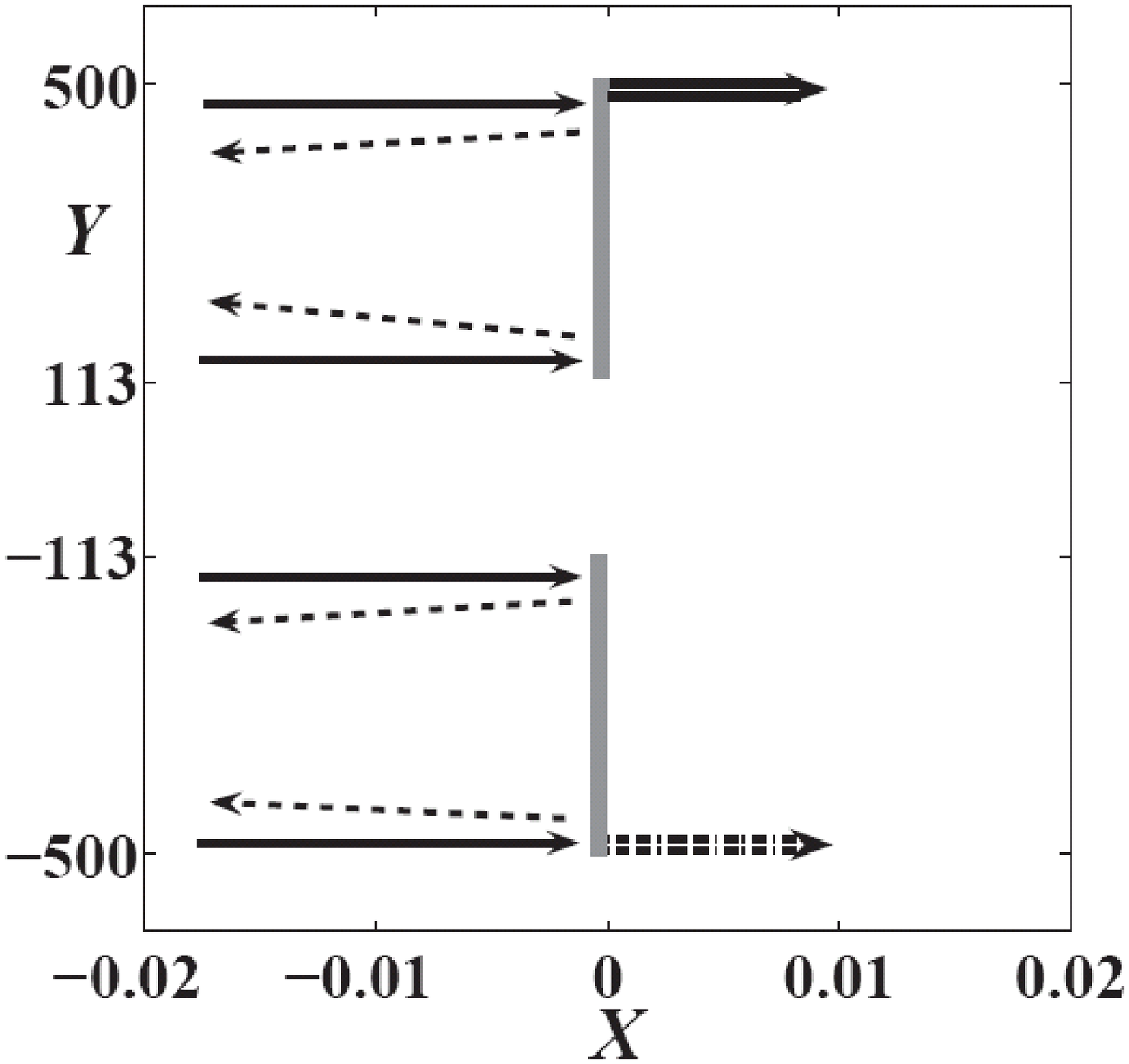}
\caption{Andreev reflection of quasiparticles from the
vortex-antivortex pair travelling in the same direction. The solid
lines denote quasiparticles travelling left to right; the dotted
lines denote the reflected quasiholes. The solid double arrow
denotes the path of the vortex, initially located at
$(Q_{x1},\,O_{y1})=(0,\,500)$ with separation $D=1000$; the
dash-dotted double arrow denotes the path of the antivortex,
initially located at $(Q_{x2},\,Q_{y2})=(0,\,-500)$. The thick
vertical grey lines denote the shadows of the vortices.}
\label{fig:8}
\end{figure}

\begin{figure}
\centering
\includegraphics[angle=0,height=2.5in]{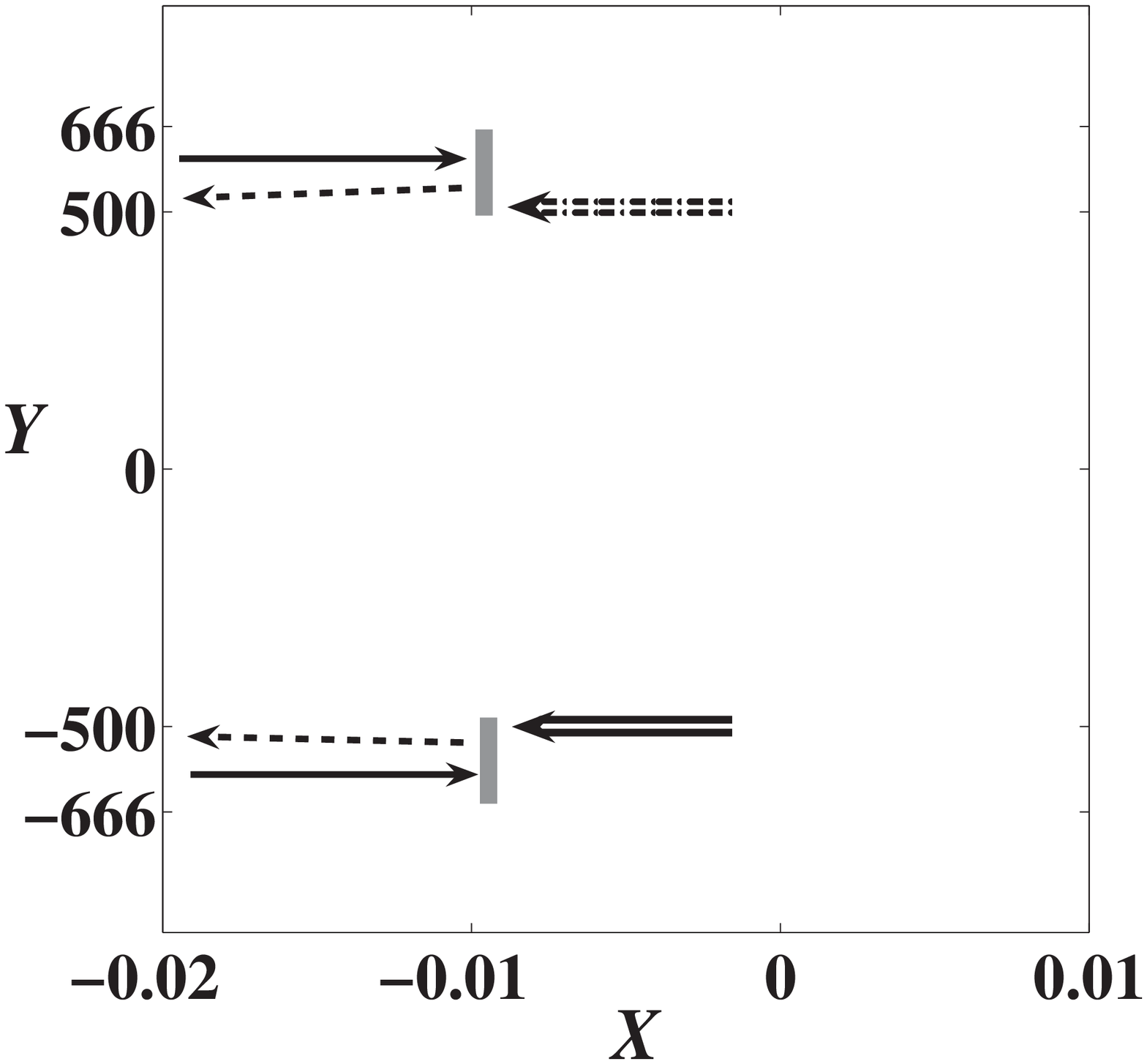}
\caption{Andreev reflection of quasiparticles from the
vortex-antivortex pair, with separation $D=1000$, travelling in the
opposite direction. The solid lines denote quasiparticles travelling
left to right; the dotted lines denote the reflected quasiholes. The
dash-dotted double arrow denotes the path of the vortex, initially
located at $(Q_{x1},\,Q_{y1})=(0,\,-500)$; the solid double arrow
denotes the path of the antivortex, initially located at
$(Q_{x2},\,Q_{y2})=(0,\,500)$. The thick vertical grey lines denote
the shadows of the vortices. Note that the shadows are much smaller
than in Fig.~\ref{fig:8}.} \label{fig:9}
\end{figure}

A vortex and an antivortex, set at distance $D$ from each other,
move through the fluid with (dimensionless) translational velocity
\begin{equation}
V_{tran}=\frac{1}{2\pi D},~~~~~D=\frac{d}{\xi_0}.
\label{eq:trans_vel_dim}
\end{equation}
The same estimates which we have made at the beginning of the
previous section apply. As before, we consider quasiparticles with
initial momentum ${\bf\Pi}_0=(1.0001,\,0)$ and initial position
$(X_0,\,Y_0)$ with $X_0=-10^4$ fixed and varying $Y_0$. Firstly we
consider the case in which the vortex-antivortex pair and the
quasiparticle move in the same direction. Let the positive
(anticlockwise) vortex be located at $(Q_{x1},\,Q_{y1})=(0,\,500)$
and the negative (clockwise) vortex be at
$(Q_{x2},\,Q_{y2})=(0,\,-500)$, with $D=1000$ the separation between
the vortices.  We find that the total shadow of the vortex
configuration is $S_1+S_2=774$, as shown in Fig.~\ref{fig:8}.
Secondly we consider the case in which the vortex-antivortex pair
and the quasiparticle move in the opposite directions, letting the
positive vortex be at $(Q_{x1},\,Q_{y1})=(0,\,-500)$ and the
negative vortex at $(Q_{x2},\,Q_{y2})=(0,\,500)$: we find that the
total shadow is greatly reduced: $S_1+S_2=332$, as shown in
Fig.~\ref{fig:9}. In both cases, during the timescale of the
calculation, the vortex pair moves by only $0.01$. We conclude that
the relative motion of vortices and excitations has a strong effect
on the shadow.

Now we examine the dependence of the Andreev reflection on $D$.
Firstly, we consider the case in which the quasiparticle and the
vortex pair move in the same direction. We said that, at
$D=1000$, the total shadow is $S_1+S_2=774$. If we reduce $D$,
the total shadow increases, and at the
value $D=940$ the two shadows merge into a single shadow of
size $S_1+S_2=940$. Upon further reduction of $D$, the total
shadow decreases; for example, when $D=100$, $S_1+S_2=122$,
and when $D=10$ we have $S_1+S_2=34$.
Secondly, we consider the case in which the quasiparticle and
the vortex pair move in opposite directions. We have said that
if $D=1000$ the total shadow is only $S_1+S_2=332$. If $D$
decreases, $S_1+S_2$ decreases too: at $D=100$ and $D=10$
we have $S_1+S_2=222$ and $88$ respectively. We conclude that,
independently of $D$, the total shadow of a vortex pair travelling
in the opposite direction of the quasiparticle is about half that
of a vortex pair travelling in the same direction.

The shadow of the vortex-antivortex pair also depends on the angle
$\alpha$ between the direction of propagation of the quasiparticle
and the direction of motion of the vortex-antivortex pair. Consider
a vortex-antivortex pair of separation $D=100$. We have already seen
that if $\alpha=0$ (vortex pair moving in the same direction as the
quasiparticle), then $S=122$. If the positive vortex is at
$(35.33,\,35.33)$ and the negative vortex is at $(-35.33,\,-35.33)$,
the angle is $\alpha=-45^o$ and the shadow increases to $168$. If
the positive vortex is at $(50,\,0)$ and the negative vortex at
$(-50,\,0)$, the angle is $\alpha=-90^o$ and the shadow increases
further to $S=233$. Finally, as said before, if the positive vortex
is at $(0,\,-50)$ and the negative vortex is at $(0,\,50)$ (the
vortex pair moving in the direction opposite to that of the
quasiparticle), then $\alpha=-180^o$ and $S=222$.

\section{Many vortices} \label{many}

We assume again that the quasiparticle has initial momentum ${\bf
\Pi}_0=(1.0001,\,0)$ and initial position $(X_0,\,Y_0)$ with
$X_0=-10^4$; we vary $Y_0$ and determine the total shadow of some
simple vortex configurations.

In the first numerical experiment we initially place five vortex
points in the square $-250 \le X \le 250$, $-250 \le Y \le 250$.
More precisely, the initial positions of the vortices are
$(-250,\,0)$, $(0,\,0)$, $(250,\,0)$, $(250,\,-250)$ and
$(-250,\,-250)$.

\begin{figure}
\centering
\includegraphics[angle=0,height=2.6in]{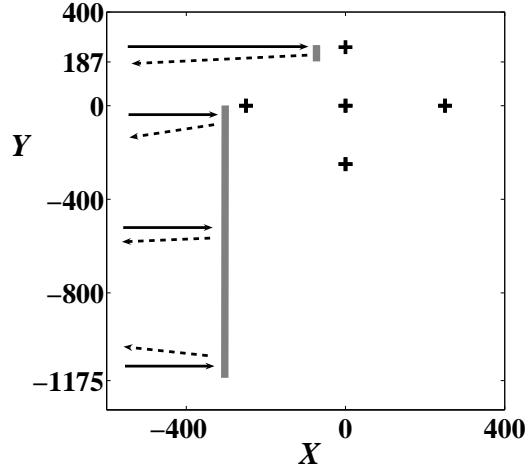}
\caption{Trajectories of quasiparticles (solid lines), of initial
momentum ${\bf\Pi}_0=(1.0001,\,0)$ and initial position
$(-10^4,\,Y_0)$ for varying $Y_0$, interacting with a configuration
of five positive vortices (denoted by crosses) initially located at
$(-250,\,0)$, $(0,\,0)$, $(250,\,0)$, $(250,\,-250)$ and
$(-250,\,-250)$. The reflected quasiholes are indicated as dotted
lines; the shadow of the vortex configuration is the thick grey
line.} \label{fig:10}
\end{figure}

If the vortices have the same (positive) polarity, they rotate
around each other, forming the 2-dimensional equivalent of a vortex
bundle of total circulation $5 \kappa$; we find (see
Fig.~\ref{fig:10}) that the total shadow of the vortex configuration
is $S=1238$, which is less than five times the shadow of five
individual vortices.

\begin{figure}
\centering
\includegraphics[angle=0,height=2.6in]{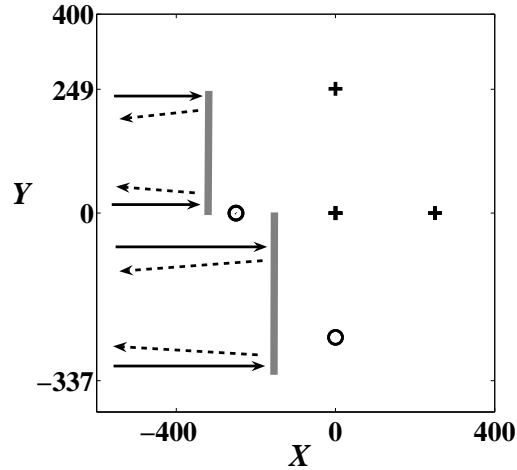}
\caption{Trajectories of quasiparticles (solid lines), of initial
momentum ${\bf\Pi}_0=(1.0001,\,0)$ and initial position
$(-10^4,\,Y_0)$ for varying $Y_0$, interacting with an initial
configuration of three positive vortices (denoted by crosses)
located at $(0,\,0)$, $(250,\,0)$ and $(0,\,250)$ and two negative
vortices located at $(-250,\,0)$ and $(0,\,-250)$ denoted by
circles. The reflected quasiholes are indicated as dotted lines; the
shadow of the vortex configuration is the thick grey line.}
\label{fig:11}
\end{figure}

\begin{figure}
%\centering \epsfig{figure=shadows_4_3.eps,height=3.5in,angle=0}
\centering \epsfig{figure=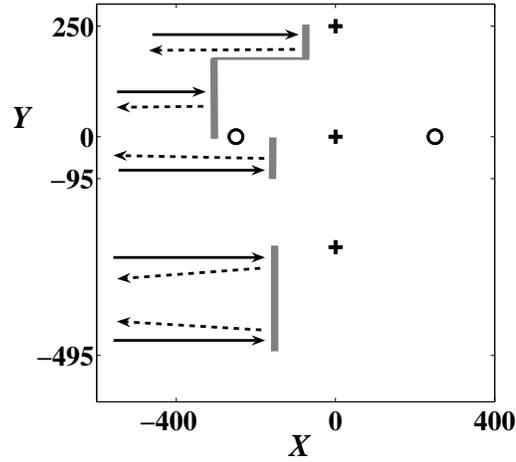,height=2.6in,angle=0}
\caption{ Trajectories of quasiparticles (solid lines), of initial
momentum ${\bf\Pi}_0=(1.0001,\,0)$ and initial position
$(-10^4,\,Y_0)$ for varying $Y_0$, interacting with an initial
configuration of three positive vortices (crosses) located at
$(0,\,0)$, $(0,\,250)$ and $(0,\,-250)$ and two negative vortices
(circles) located at $(250,\,0)$ and $(-250,\,0)$. The reflected
quasiholes are indicated as dotted lines; the total shadow of the
vortex configuration,
%$250+95+(495-251)=589$.
$S=589$, is the thick grey line. Note the gap in the shadow within
the interval $-95\leq Y_0\leq 0$.} \label{fig:12}
\end{figure}

If we change the sign of some of the vortices, the total shadow
which is cast changes dramatically. For example, let the vortices at
$(0,\,0)$, $(0,\,250)$ and $(250,\,0)$ be positive, and the vortices
at $(-250,\,0)$ and $(0,\,-250)$ be negative. The net circulation is
now $\kappa$ and the total shadow is $S=585$ (see
Fig.~\ref{fig:11}), which is much less than the previous value, but
still more than the shadow of a single isolated vortex. A similar
value of the total shadow, $S=589$, is obtained if the positive
vortices are at $(0,\,0)$, $(0,\,250)$, $(0,\,-250)$ and the
negative vortices are at $(-250,\,0)$, $(250,\,0)$, which
corresponds to the same total circulation $\kappa$ as before, see
Fig.~\ref{fig:12}.

\begin{figure}
%\centering \epsfig{figure=shadows_4_4.eps,height=3.5in,angle=0}
\centering \epsfig{figure=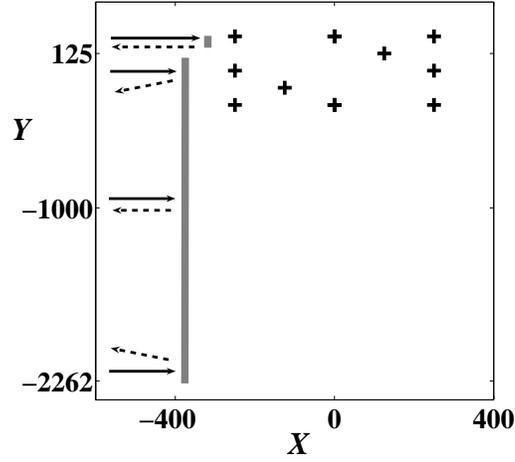,height=2.6in,angle=0}
\caption{ Schematic trajectories of quasiparticles (solid arrows),
of initial momentum ${\bf\Pi}_0=(1.0001,\,0)$ and initial position
$(-10^4,\,Y_0)$ for varying $Y_0$ interacting with an initial
configuration of ten positive vortices (crosses) located at
$(-250,\,0)$, $(-250,\,250)$, $(-125,\,-125)$, $(0,\,-250)$,
$(0,\,250)$, $(125,\,125)$, $(250,\,-250)$, $(250,\,0)$ and
$(250,\,250)$. The reflected quasiholes are indicated as dashed
arrows. The thick grey vertical lines indicate the shadows of the
vortices. The total shadow is $S=2445$.} \label{fig:13}
\end{figure}

\begin{figure}
%\centering \epsfig{figure=shadows_4_5.eps,height=3.5in,angle=0}
\centering \epsfig{figure=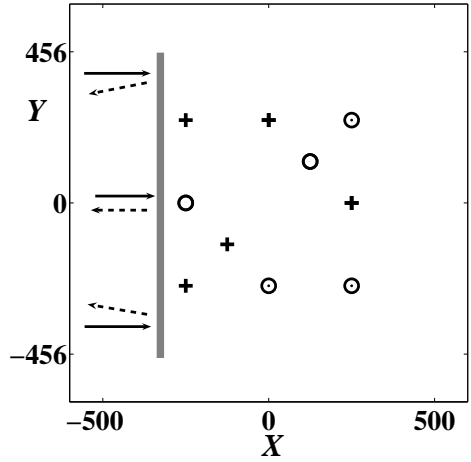,height=2.6in,angle=0}
\caption{ Schematic trajectories of quasiparticles (solid arrows),
of initial momentum ${\bf\Pi}_0=(1.0001,\,0)$ and initial position
$(-10^4,\,Y_0)$ for varying $Y_0$, interacting with an initial
configuration of five positive vortices (crosses) located at
$(-250,\,-250)$, $(-250,\,250)$, $(-125,\,-125)$, $(0,\,250)$, and
five antivortices (circles) at the $(-250,\,0)$, $(0,\,-250)$,
$(125,\,125)$, $(250,\,-250)$ and $(250,\,250)$. The reflected
quasiholes are indicated as dashed arrows. The thick grey vertical
line indicates the shadow of the vortices. The total shadow is
$S=902$.} \label{fig:14}
\end{figure}

\begin{figure}
%\centering \epsfig{figure=shadows_4_6.eps,height=3.5in,angle=0}
\centering \epsfig{figure=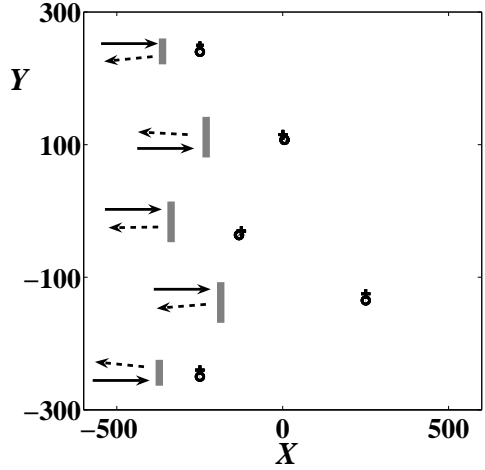,height=2.6in,angle=0}
\caption{ Schematic trajectories of quasiparticles (solid arrows),
of initial momentum ${\bf\Pi}_0=(1.0001,\,0)$ and initial position
$(-10^4,\,Y_0)$ for varying $Y_0$, interacting with an initial
configuration of five vortex-antivortex pairs located at: vortex at
$(-250,\,-240)$ and antivortex at $(-250,\,-250)$, vortex at
$(-250,\,250)$ and antivortex at $(-250,\,-240)$, vortex at
$(-125,\,-30)$ and antivortex at $(-132,\,-36)$, vortex at
$(0,\,115)$ and antivortex at $(5,\,107)$, vortex at $(250,\,-125)$
and antivortex at $(250,\,-135)$. (Note that the distance between
the vortex and the antivortex in each pair is 10 non-dimensional
units and cannot be distinguished on this figure.) The schematic
reflected quasiholes are indicated as dashed arrows. The thick grey
vertical lines indicate the shadows of the vortices. The total
shadow is $S=209$.} \label{fig:15}
\end{figure}

In the second numerical experiment we increase the vortex
density and place ten vortices (twice as many as before) in the same
square $-250 \le X \le 250$, $-250 \le Y \le 250$. We consider
three cases: (i) ten vortices of the same polarity (total
circulation $10 \kappa$, see Fig.~\ref{fig:13}), which
yields the total shadow $S=2445$ (about twice the shadow
of a bundle of five vortices); (ii) five vortices and five
antivortices (total circulation is zero, see Fig.~\ref{fig:14}),
which yields $S=902$; (iii) five vortex-antivortex pairs (total
circulation zero, see Fig.~\ref{fig:15}), which yields
$S=209$.

\begin{figure}[t]
\begin{tabular}[b]{cc}
\includegraphics[height=0.38\linewidth]{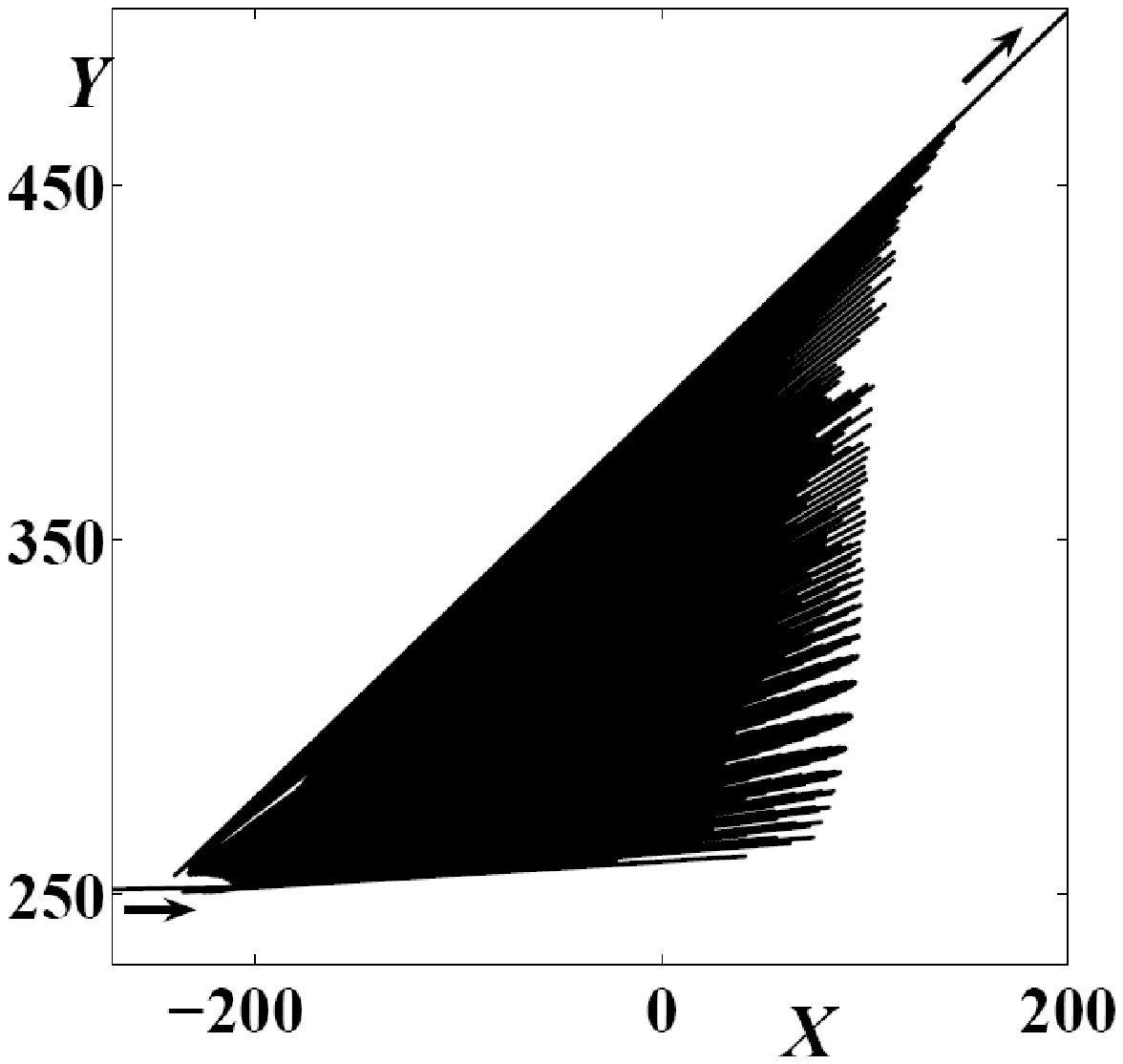}&
\includegraphics[height=0.38\linewidth]{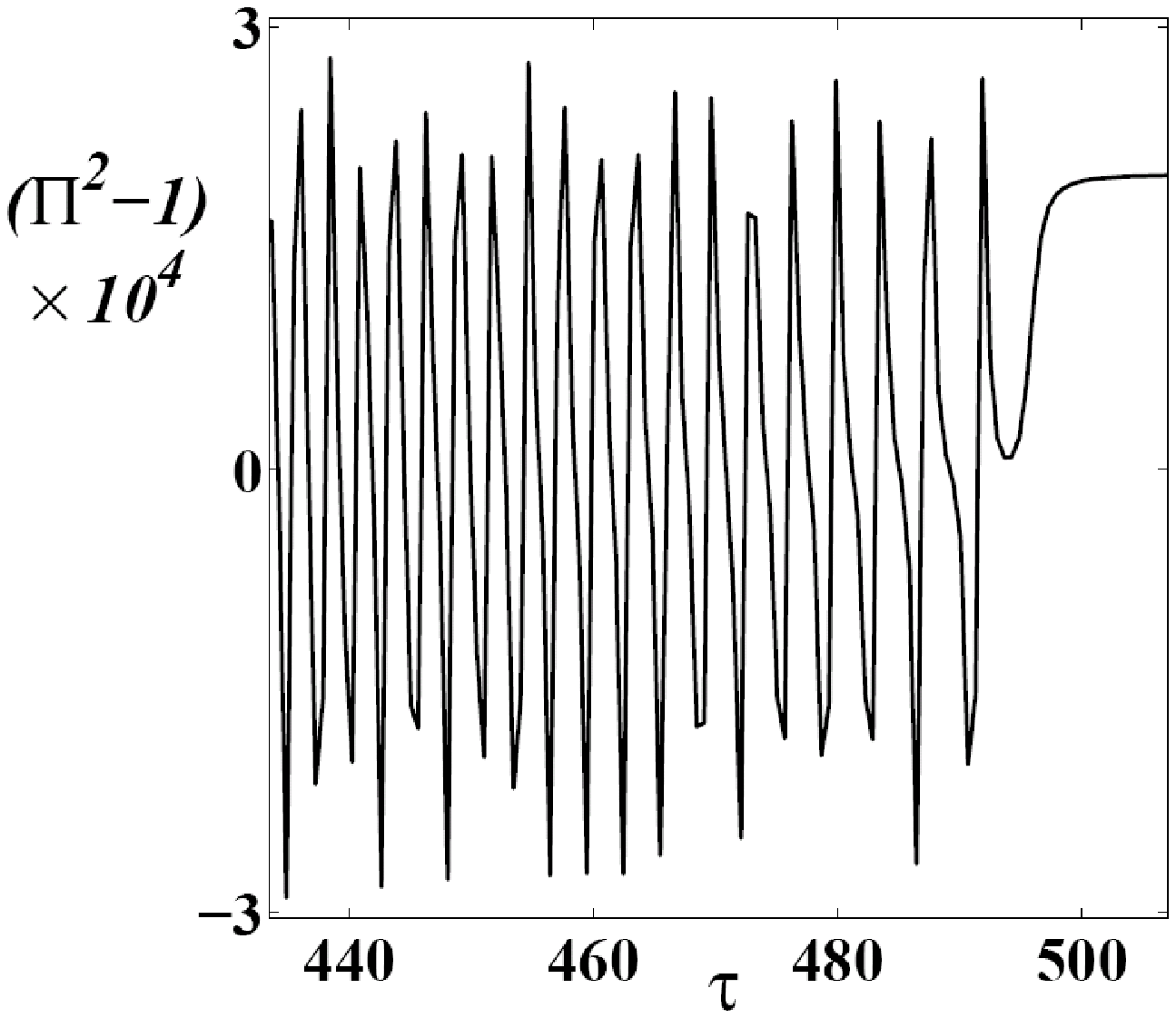}\\
\end{tabular}
\caption{Left: trajectory of the quasiparticle with initial momentum
${\bf \Pi}_0=(1.0001,\,0)$ and position $(X_0,\,Y_0)=(-10^4,\,10)$
in the presence of five positive vortices and five negative vortices
as in Fig.~\ref{fig:14}. Note the multiple reflections before the
particle's escape from the vortex region. The arrows indicate the
direction of motion. Right: plot of $\Pi^2-1$ vs time $\tau$
corresponding to a small time interval just before the escape. Note
that the quasiparticle turns into a quasihole many times, before
escaping as a quasiparticle.} \label{fig:16}
\end{figure}

We have found that it is possible that, upon impinging on complex
vortex configurations, quasiparticles experience multiple
reflections, which can be classical, Andreev, or both. An example of
multiple Andreev reflection is shown in Fig.~\ref{fig:16}.

\section{Conclusions} \label{Conclusions}

In conclusion, the above numerical experiments with simple vortex
configurations show that partial screening takes place. The total
Andreev shadow of a vortex system is not necessarily the sum of the
shadows of individual vortices, and depends not only on the distance
but also on the relative orientation between quasiparticles and
vortex motion. This does not mean that the interpretation given to
recent experiments is incorrect. It is possible that, for a large,
random vortex system, the partial screening effects which we have
found average out. If this is the case, screening effects can be
taken into account by introducing a prefactor probably of order one
for the total shadow, hence for the vortex line density which is
inferred. Numerical investigations in 3-dimensions with realistic
vortex line density are needed.

How random are vortex configurations of current experiments?
Probably only the recently discovered \cite{Golov2008,Walmsley}
ultraquantum regime is truly random. Homogeneous isotropic
turbulence contains coherent vortex structures
\cite{Farge,Sultan,Kivotides} and is organized in scales with
different energy per scale. On the other hand, provided we are
interested only in large scale properties averaged over a large
region, such as the total vortex length, the partial screening
effects can be accounted as said above.

The situation is very different when we move to rotating
turbulence \cite{Tsubota,Eltsov2008,Eltsov2006} and inhomogeneous
turbulence, particularly if there are turbulent fronts. In these cases
there is large scale anisotropy, and the Andreev reflection technique
must be used with more care than we used to. The good news is that,
by combining Andreev reflection measurements in different directions
and numerical calculations such as those we have presented, it should
be possible to gain more information about the geometry and the
anisotropy of the turbulence.

Our results also indicate that the problem of interaction between
rotons and quantized vortices in $^4$He \cite{Samuels}, leading to
calculation of the mutual friction, should be reconsidered in view
of possible screening effects analyzed above in Secs.~\ref{pair} and
\ref{antipair}.

\section{Acknowledgments}

N.S. was supported by the Georgian National Science Foundation Grant
No.~GNSF/ST06/4-018. We are grateful to S.~N.~Fisher for stimulating
discussions.

\vfill
\eject

\newpage

%%%%%%%%%%%%%%%%%%%%%%%%%%%%%%%%%%%%%%%%%%%%%%%%%%%%%%%%%%%%%%%
% REFERENCES

\newpage

%%%%%%%%%%%%%%%%%%%%%%%%%%%%%%%%%%%%%%%%%%%%%%%%%%%%%%%%%%%%%%%
% FIGURES
%%%%%%%%%%%%%%%%%%%%%%%%%%%%%%%%%%%%%%%%%%%%%%%%%%%%%%%%%%%%%%%
% FIG 1

\newpage
\vfill
\eject

%%%%%%%%%%%%%%%%%%%%%%%%%%%%%%%%%%%%%%%%%%%%%%%%%%%%%%%%%%%%%%%%%%%
% FIG 2

\newpage

\vfill
\eject

%%%%%%%%%%%%%%%%%%%%%%%%%%%%%%%%%%%%%%%%%%%%%%%%%%%%%%%%%%%%%%%%%
% FIG 3

\newpage

\vfill
\eject

%%%%%%%%%%%%%%%%%%%%%%%%%%%%%%%%%%%%%%%%%%%%%%%%%%%%%%%%%%%%%%%%%%%%
% FIG 4

\newpage

\vfill
\eject

%%%%%%%%%%%%%%%%%%%%%%%%%%%%%%%%%%%%%%%%%%%%%%%%%%%%%%%%%%%%%%%%%%
% FIG 5

\newpage

\vfill
\eject

%%%%%%%%%%%%%%%%%%%%%%%%%%%%%%%%%%%%%%%%%%%%%%%%%%%%%%%%%%%%%%
% FIG 6

\newpage

\vfill
\eject

%%%%%%%%%%%%%%%%%%%%%%%%%%%%%%%%%%%%%%%%%%%%%%%%%%%%%%%%%%%%%%%%%
% FIG 7

\newpage

\vfill
\eject

%%%%%%%%%%%%%%%%%%%%%%%%%%%%%%%%%%%%%%%%%%%%%%%%%%%%%%%%%%%%%%%%%%%
% FIG 8

\newpage

\vfill
\eject

%%%%%%%%%%%%%%%%%%%%%%%%%%%%%%%%%%%%%%%%%%%%%%%%%%%%%%%%%%%%%%
% FIG 9

\newpage

\vfill
\eject

%%%%%%%%%%%%%%%%%%%%%%%%%%%%%%%%%%%%%%%%%%%%%%%%%%%%%%%%%%%%%%%%%%
% FIG 10

\newpage

\vfill
\eject

%%%%%%%%%%%%%%%%%%%%%%%%%%%%%%%%%%%%%%%%%%%%%%%%%%%%%%%%%%%%%%%%%%%
% FIG 11

\newpage

\vfill
\eject

%%%%%%%%%%%%%%%%%%%%%%%%%%%%%%%%%%%%%%%%%%%%%%%%%%%%%%%%%%%%%%%%%
% FIG 12

\newpage

\vfill
\eject

%%%%%%%%%%%%%%%%%%%%%%%%%%%%%%%%%%%%%%%%%%%%%%%%%%%%%%%%%%%%%%%%%%
% FIG 13

\newpage

\vfill
\eject

%%%%%%%%%%%%%%%%%%%%%%%%%%%%%%%%%%%%%%%%%%%%%%%%%%%%%%%%%%%%%%%%
% FIG 14

\newpage

\vfill
\eject

%%%%%%%%%%%%%%%%%%%%%%%%%%%%%%%%%%%%%%%%%%%%%%%%%%%%%%%%%%%%%%%%%
% FIG 15

\newpage

\vfill
\eject

%%%%%%%%%%%%%%%%%%%%%%%%%%%%%%%%%%%%%%%%%%%%%%%%%%%%%%%%%%%%%%%%%%%%%
% FIG 16

\vfill
\eject

\end{document}